\documentclass[prl,twocolumn,superscriptaddress]{revtex4-2}

\usepackage{amsmath}
\usepackage{amssymb}
\usepackage{graphicx}
\usepackage{bbm}
\usepackage{wasysym}
\usepackage{newtxtext}
\usepackage[varvw]{newtxmath}
\usepackage{hyperref}

\hypersetup{
    colorlinks,
    linkcolor={blue},
    citecolor={blue},
    urlcolor={blue}
}

\setcitestyle{notesep={}}

\graphicspath{{./}{./figs/}}

\newcommand{\NCTO}{Na\textsubscript{2}Co\textsubscript{2}TeO\textsubscript{6}}
\newcommand{\NCSO}{Na\textsubscript{3}Co\textsubscript{2}SbO\textsubscript{6}}
\newcommand{\BCAO}{BaCo\textsubscript{2}(AsO\textsubscript{4})\textsubscript{2}}

\newcommand{\llangle}{\langle\!\langle}
\newcommand{\rrangle}{\rangle\!\rangle}
\newcommand{\lllangle}{\langle\!\langle\!\langle}
\newcommand{\rrrangle}{\rangle\!\rangle\!\rangle}

\DeclareMathAlphabet{\mymathbb}{U}{BOONDOX-ds}{m}{n}

\begin{document}

\title{Triple-q order in Na\textsubscript{2}Co\textsubscript{2}TeO\textsubscript{6} from proximity to hidden-SU(2)-symmetric point}

\author{Wilhelm G.\ F.\ Kr\"uger}
\thanks{WGFK and WC contributed equally to this work.}
\affiliation{Institut f\"ur Theoretische Physik and W\"urzburg-Dresden Cluster of Excellence ct.qmat, TU Dresden, 01062 Dresden, Germany}

\author{Wenjie Chen}
\thanks{WGFK and WC contributed equally to this work.}
\affiliation{International Center for Quantum Materials, School of Physics, Peking University, Beijing 100871, China}

\author{Xianghong Jin}
\affiliation{International Center for Quantum Materials, School of Physics, Peking University, Beijing 100871, China}

\author{Yuan Li}
\affiliation{International Center for Quantum Materials, School of Physics, Peking University, Beijing 100871, China}
\affiliation{Collaborative Innovation Center of Quantum Matter, Beijing 100871, China}

\author{Lukas Janssen}
\affiliation{Institut f\"ur Theoretische Physik and W\"urzburg-Dresden Cluster of Excellence ct.qmat, TU Dresden, 01062 Dresden, Germany}

\begin{abstract}
In extended Heisenberg-Kitaev-Gamma-type spin models, hidden-SU(2)-symmetric points are isolated points in parameter space that can be mapped to pure Heisenberg models via nontrivial duality transformations.
Such points generically feature quantum degeneracy between conventional single-$\mathbf q$ and exotic multi-$\mathbf q$ states.
We argue that recent single-crystal inelastic neutron scattering data place the honeycomb magnet \NCTO\ in proximity to such a hidden-SU(2)-symmetric point.
The low-temperature order is identified as a triple-$\mathbf q$ state arising from the N\'eel antiferromagnet with staggered magnetization in the out-of-plane direction via a 4-sublattice duality transformation.
This state naturally explains various distinctive features of the magnetic excitation spectrum, including its surprisingly high symmetry and the dispersive low-energy and flat high-energy bands.
Our result demonstrates the importance of bond-dependent exchange interactions in cobaltates, and illustrates the intriguing magnetic behavior resulting from them.
\end{abstract}

\date{\today}

\maketitle

\paragraph{Introduction.}
%
In the search for quantum spin liquids, the Kitaev honeycomb model~\cite{kitaev06} plays a pivotal role.
It features bond-dependent exchange interactions between neighboring sites and may be realizable in magnetic Mott insulators involving heavy ions~\cite{trebst21}.
These materials are characterized by entangled spin and orbital degrees of freedom. The physical spin is no longer a good quantum number and the pseudospin degree of freedom replacing it typically features a lower symmetry that involves combined rotations in both pseudospin and lattice spaces.
On the honeycomb lattice, for instance, spin-orbit coupling reduces the continuous Heisenberg SU(2) spin symmetry down to a discrete $C_3^\ast$ symmetry of $2\pi/3$ pseudospin rotations about a high-symmetry axis, combined with corresponding lattice rotations~\cite{janssen19}.
This leads to a large number of symmetry-allowed exchange couplings, including Kitaev and off-diagonal $\Gamma$ and $\Gamma'$ interactions~\cite{chaloupka13, rau14, winter17}.
Within this large parameter space, there exist, however, isolated points that feature a \emph{hidden} continuous SU(2) symmetry that is different from the usual Heisenberg symmetry~\cite{chaloupka10, kimchi14}. These hidden-SU(2)-symmetric points can be mapped to standard Heisenberg points via duality transformations involving local pseudospin rotations~\cite{chaloupka15}.
The hidden SU(2) symmetry inevitably causes an SU(2) degeneracy of the model's quantum ground state. Perturbations away from such hidden-SU(2)-symmetric points can then stabilize a variety of different states, including states characterized by multiple ordering wavevectors (multi-$\mathbf q$ states), noncollinear pseudospin configurations, and/or large magnetic unit cells~\cite{chaloupka15, janssen16}.
An interesting such state that has been studied in theoretical works of Heisenberg-Kitaev models on the honeycomb lattice is a noncollinear triple-$\mathbf q$ state in which each domain features three Bragg peaks in the first crystallographic Brillouin zone, with an 8-site magnetic unit cell~\cite{janssen16, consoli20}.

\begin{figure*}[t]
\includegraphics[width=\linewidth]{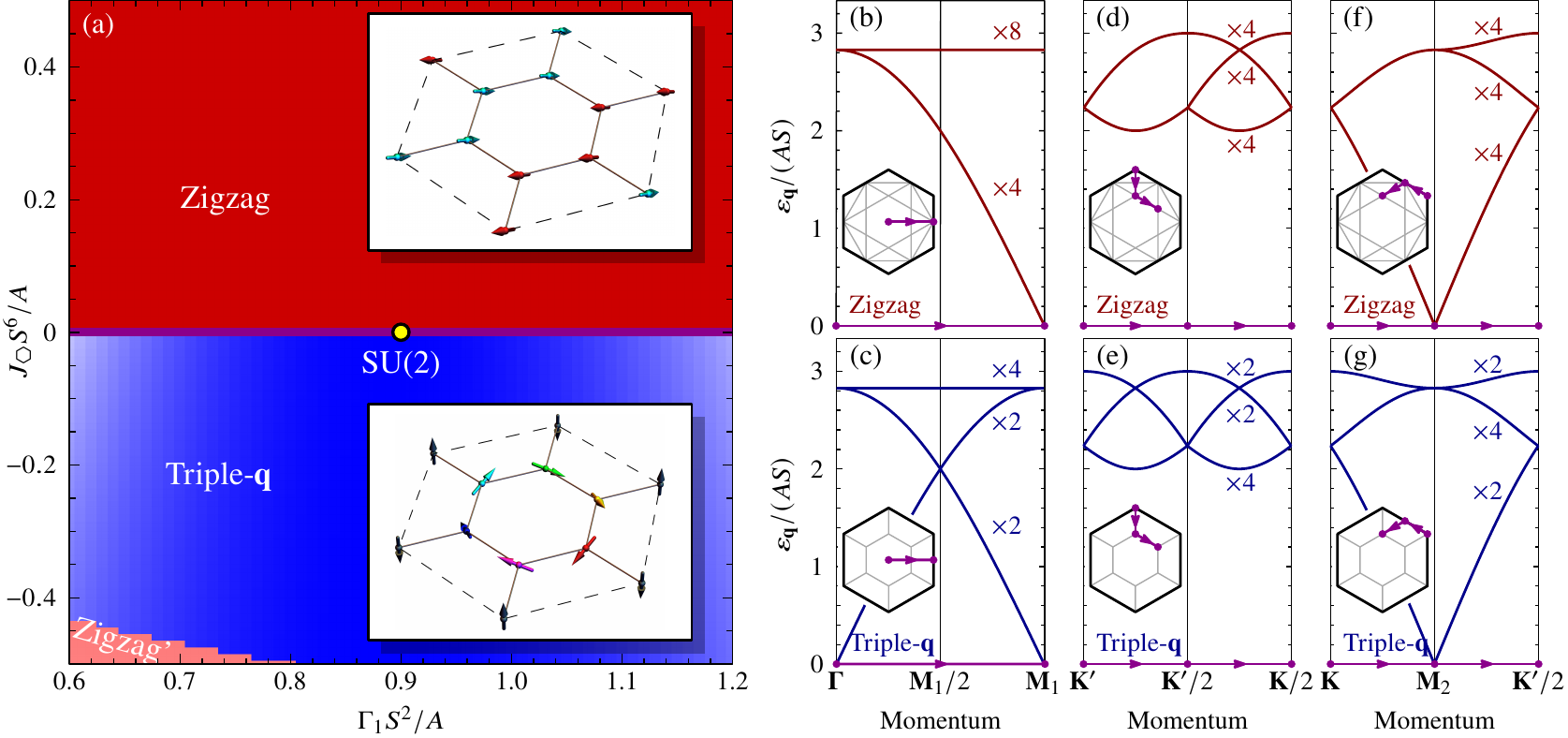}
\caption{%
(a)~Classical phase diagram of HK$\Gamma\Gamma'$ model in vicinity of hidden-SU(2)-symmetric point (yellow dot)
in the plane spanned by off-diagonal coupling $\Gamma_1$ and ring exchange coupling $J_{\hexagon}$, from classical Monte Carlo simulations.
For $J_{\hexagon}>0$ ($J_{\hexagon}<0$), zigzag (triple-$\mathbf q$) order with a 4-site (8-site) magnetic unit cell is stabilized, see insets. Color scales indicate Bragg-peak intensities in different phases in arbitrary units. 
The zigzag' phase at small $\Gamma_1$ and large negative $J_{\hexagon}$ is characterized by a 4-site unit cell, in which the spins along zigzag chains become slightly noncollinear.
(b--g)~Magnon dispersions from linear spin-wave theory at hidden-SU(2)-symmetric point for (b,d,f) domain-averaged zigzag order and (c,e,g) triple-$\mathbf q$ order, along different trajectories within the first structural Brillouin zone, as displayed in insets. The triple-$\mathbf q$ spectra are symmetric along these trajectories, while the zigzag spectra are not. This can be understood as a consequence of the difference in shape and size of the magnetic Brillouin zones, indicated in gray in insets.}
\label{fig:spectrum-SU2}
\end{figure*}

Identifying such a multi-$\mathbf q$ state in an experiment, however, is a challenging task. 
The structure of the duality transformation limits the position of the Bragg peaks to a fixed set of symmetry-related points in the Brillouin zone. The Bragg-peak pattern of a single magnetic domain of a multi-$\mathbf q$ state, as measurable, for instance, in neutron diffraction experiments, is therefore identical to the averaged pattern of multiple domains of a corresponding single-$\mathbf q$ state.
Since domain mixing effects are difficult to exclude in most experimental setups, the diffraction pattern of a multi-$\mathbf q$ state can be easily misinterpreted as a domain-averaged single-$\mathbf q$ pattern.
A class of materials exemplifying this difficulty are $d^7$ honeycomb cobaltates, such as \NCTO, \NCSO, or \BCAO. These compounds have received significant interest recently as possible realizations of the Kitaev honeycomb model~\cite{liu18, sano18, yan19, liu20, yao20, songvilay20, lin21, chen21, lee21, hong21, samarakoon21, kim22, mukherjee22, sanders22, yang22, yao22a, li22, zhong20, shi21, zhang22, winter22}.
All three mentioned examples exhibit long-range magnetic order at low temperatures. Despite various experimental efforts, however, the natures of the magnetic ground states have remained under debate. In \NCTO, for instance, many experimental observations resemble those of $\alpha$-RuCl$_3$~\cite{janssen19, takagi19}, which has led to the assumption that the ground state features the same zigzag type of single-$\mathbf q$ order~\cite{lefrancois16, bera17}. Recent inelastic neutron scattering results on high-quality single crystals, however, have revealed a single lowest-energy magnon branch without any sign of domain superposition~\cite{chen21}, and at least five additional nonoverlapping spin-wave branches at higher energy~\cite{yao22a}, which may be indicative of a multi-$\mathbf q$ state.
Furthermore, nuclear magnetic resonance experiments have suggested a noncollinear magnetic structure that is inconsistent with the simple zigzag ordering~\cite{lee21}.
These results are at odds with all spin models proposed for \NCTO\ to date, all of which featuring a zigzag ground state~\cite{songvilay20, lin21, kim22, sanders22}. 
A similar ambiguity between single-$\mathbf q$ and multi-$\mathbf q$ scenarios arises for \NCSO~\cite{li22} on the honeycomb lattice, as well as for Co$_{1/3}$TaS$_2$ on the triangular lattice~\cite{park23}.

In this work, we demonstrate how a multi-$\mathbf q$ state that features Bragg peaks at symmetry-related points in the crystallographic Brillouin zone can be uniquely identified. 
We exploit the fact that the \emph{shape} of the magnetic Brillouin zone in a multi-$\mathbf q$ state differs from those of the corresponding single-$\mathbf q$ states.
This leads to different symmetries of the magnetic excitation energies.
Applying this approach to \NCTO, for which high-resolution spectra are available~\cite{yao22a}, reveals a noncollinear triple-$\mathbf q$ ground state with a hexagonally-shaped magnetic unit cell featuring 8 magnetic sites.
An effective model, which features nonbilinear exchange interactions in addition to the usual Heisenberg-Kitaev-Gamma-type exchanges, stabilizes the triple-$\mathbf q$ order and reproduces important features in the measured magnetic excitation spectrum very well.

\paragraph{Model.}
%
To understand the novel physics of the $d^7$ honeycomb cobaltates, consider an extended Heisenberg-Kitaev-Gamma-Gamma$'$ (HK$\Gamma\Gamma'$) model,
\begin{align} \label{eq:model}
	\mathcal H
	& = \sum_{\gamma = x,y,z} \sum_{\langle ij \rangle_\gamma} \Bigl[J_1 \mathbf S_i \cdot \mathbf S_j + K_1 S_i^\gamma S_j^\gamma + \Gamma_1 (S^\alpha_i S^\beta_j + S^\beta_i S^\alpha_j) 
    \nonumber \displaybreak[1] \\ & \quad
    + \Gamma'_1 (S^\gamma_i S^\alpha_j + S^\alpha_i S^\gamma_j + S^\gamma_i S^\beta_j + S^\beta_i S^\gamma_j) \Bigr]
    + \sum_{\lllangle ij \rrrangle} J_{3} \mathbf S_i \cdot \mathbf S_j
    \nonumber \displaybreak[1] \\ & \quad
    + \sum_{\llangle ij \rrangle^A} J_{2}^A \mathbf S_i \cdot \mathbf S_j
    + \sum_{\llangle ij \rrangle^B} J_{2}^B \mathbf S_i \cdot \mathbf S_j
    + \mathcal H_\text{nbl}\,,
\end{align}
where $\langle ij\rangle_\gamma$ denotes first-neighbor $\gamma$ bonds on the honeycomb lattice, and $\lllangle ij \rrrangle$ are third-neighbor bonds along opposite sites of the same hexagon.
$J_1$ and $J_3$ are the corresponding first- and third-neighbor Heisenberg couplings, $K_1$ the first-neighbor Kitaev coupling, and $\Gamma_1$ and $\Gamma_1'$ the two first-neighbor symmetric off-diagonal couplings.
$\llangle ij \rrangle^A$ and $\llangle ij \rrangle^B$ correspond to second-neighbor bonds between sites on the same sublattice $A$ and $B$, respectively.
Here, we have taken a possible sublattice symmetry breaking into account, parametrized by the two second-neighbor Heisenberg couplings $J_2^A$ and $J_2^B$. In \NCTO, sublattice symmetry breaking arises from the two crystallographically inequivalent Co$^{2+}$ sites~\cite{yao20}.
Finally, $\mathcal H_\text{nbl}$ denotes nonbilinear exchange interactions, such as ring exchanges.
These can be understood as leading corrections to the first-neighbor bilinear exchange in the strong-coupling expansion of the Hubbard model~\cite{takahashi77, macdonald88, yang10, yang12}, and are important in a number of $3d$ materials, including various chromium-, manganese-, and copper-based magnets~\cite{kvashnin20, fedorova15, dallapiazza12, larsen19}.
They will be specified below.

\begin{figure*}[t]
\includegraphics[width=\linewidth]{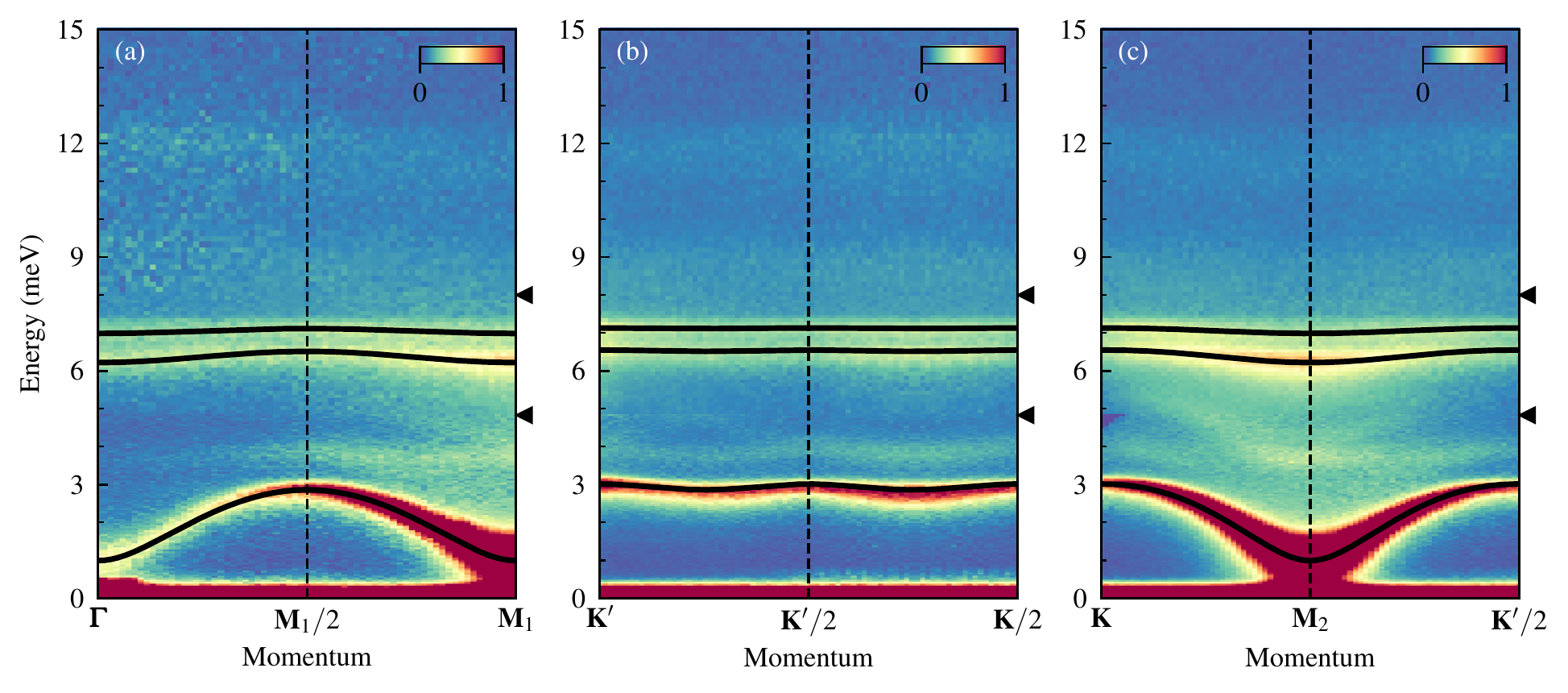}
\caption{Inelastic neutron scattering intensity of \NCTO\ at $T = 5~\mathrm{K}$ along momentum trajectories indicated in insets of Fig.~\ref{fig:spectrum-SU2}(b--g). Data are measured with $E_i = 19.4~\mathrm{meV}$ (upper part), $E_i = 10.0~\mathrm{meV}$ (middle part), $E_i = 6.1~\mathrm{meV}$ (lower part), and merged at energies indicated by left-pointing triangles. All visible magnon bands are symmetric along these trajectories, indicating a hexagonal magnetic Brillouin zone with an 8-site real-space magnetic unit cell. Black solid lines indicate three representative magnon band dispersions parameterized with phenomenological functions~\cite[(d)]{supplemental}, which are our fitting targets.}
\label{fig:spectrum-NCTO}
\end{figure*}

\paragraph{Hidden SU(2) symmetries.}

For particular values of the exchange couplings, the HK$\Gamma\Gamma'$ model defined above features hidden SU(2) symmetries that can be revealed via suitable duality transformations~\cite{chaloupka15}.
These transformations correspond to local spin rotations and map the hidden-SU(2)-symmetric HK$\Gamma\Gamma'$ model for the original spins to a pure Heisenberg model for the dual spins.
In the parameter regime $K_1 < 0$, relevant for the cobaltates~\cite{liu18, sano18}, such a hidden-SU(2)-symmetric point can be identified by a $\mathcal T_1 \mathcal T_4$ transformation. Here, $\mathcal T_1$ corresponds to a global $\pi$ rotation about the $[111]$ axis in cubic spin space, whereas $\mathcal T_4$ denotes the $4$-sublattice transformation that corresponds to a $\pi$ rotation about $x$, $y$, and $z$ axes at sublattices $1$, $2$, and $3$, respectively, and no rotation at sublattice~$4$ \cite[(a)]{supplemental}.
Inverting the transformation maps the HK$\Gamma\Gamma'$ model with $\mathcal H_\text{nbl} = 0$ and bilinear couplings
\begin{align} \label{eq:SU(2)}
    (J_1, K_1, \Gamma_1, \Gamma_1', J_2^A, J_2^B)_\text{SU(2)} = (-\tfrac19, -\tfrac23, \tfrac89, -\tfrac49, 0, 0)A\,,
\end{align}
and arbitrary $J_3$, 
where $A>0$ corresponds to the overall energy scale,
to a $\tilde J_1$-$\tilde J_3$ Heisenberg model for the dual spins, with dual couplings $\tilde J_1 = A$ and $\tilde J_3 = J_3$.
At this hidden-SU(2)-symmetric point, the HK$\Gamma\Gamma'$ model features an SU(2)-degenerate ground-state manifold, each member of which can be mapped to an associated ground state of the dual $\tilde J_1$-$\tilde J_3$ model, and vice versa.
The N\'eel state in the dual model with staggered magnetization along the $x$, $y$, and $z$ cubic axes, for instance, maps to the zigzag state in the original model with antiferromagnetic neighbor pairs along $x$, $y$, and $z$ bonds, respectively, see upper inset of Fig.~\ref{fig:spectrum-SU2}(a).
The N\'eel state with staggered magnetization along the $[111]$ axis, by contrast, maps to a triple-$\mathbf q$ state with an $8$-site magnetic unit cell and a vortex spin structure on 1/4 of the hexagonal plaquettes, see lower inset of Fig.~\ref{fig:spectrum-SU2}(a).

While bilinear exchange perturbations cannot lift the SU(2) degeneracy in the classical limit, and quantum fluctuations typically favor collinear zigzag states~\cite[(b)]{supplemental}, a triple-$\mathbf q$ state can be stabilized by nonbilinear exchange interactions.
To be specific, let us consider the ring exchange that is generated in the strong-coupling expansion of the single-band $t$-$U$ Hubbard model on the honeycomb lattice~\cite{yang12},
\begin{align} \label{eq:ring}
&\mathcal H_\text{nbl} = \frac{J_{\hexagon}}{6} \sum_{\langle ijklmn \rangle } \Bigl[
2(\mathbf S_i \cdot \mathbf S_j) (\mathbf S_k \cdot \mathbf S_l) (\mathbf S_m \cdot \mathbf S_n)
\nonumber \allowdisplaybreaks[1] \\ & \
- 6(\mathbf S_i \cdot \mathbf S_k) (\mathbf S_j \cdot \mathbf S_l) (\mathbf S_m \cdot \mathbf S_n)
+ 3 (\mathbf S_i \cdot \mathbf S_l) (\mathbf S_j \cdot \mathbf S_k) (\mathbf S_m \cdot \mathbf S_n)
\nonumber \allowdisplaybreaks[1] \\ & \
+ 3 (\mathbf S_i \cdot \mathbf S_k) (\mathbf S_j \cdot \mathbf S_m) (\mathbf S_l \cdot \mathbf S_n)
- (\mathbf S_i \cdot \mathbf S_l) (\mathbf S_j \cdot \mathbf S_m) (\mathbf S_k \cdot \mathbf S_n)
\nonumber \allowdisplaybreaks[1] \\ & \
+ \text{cyclic permutations of $(ijklmn)$}
\Bigr]\,,
\end{align}
with summation over elemental plaquettes involving sites $(ijklmn)$ in counter-clockwise order.
In \NCTO, the above nonbilinear interaction arises from virtual ring exchange processes of cobalt electrons around elemental plaquettes on the honeycomb lattice, with the prefactors originating from the number of symmetry-equivalent processes and the involved permutations~\cite{godfrin88}.
Formally, this form of the interaction is generated at sixth order in the strong-coupling expansion. At intermediate coupling, however, it represents the main correction to the nearest-neighbor exchange $J_1$, exceeding the second-neighbor exchange $J_2$, which formally arises already at fourth order in the expansion: For instance, at $t/U = 0.2$, Ref.~\cite{yang12} finds $|J_{\hexagon}/J_2| \simeq 4$.
Near the hidden-SU(2)-symmetric point, this ring exchange favors triple-$\mathbf q$ order for $J_{\hexagon} < 0$, see Fig.~\ref{fig:spectrum-SU2}(a).
For the cobaltates, various different nonbilinear exchange terms are in principle possible, and the ring coupling $J_{\hexagon}$ defined above constitutes only one representative axis in a large parameter space.
For symmetry reasons, however, all of these couplings will generically have the same qualitative effect: They will favor zigzag order on one side of the hidden-SU(2)-symmetric point and triple-$\mathbf q$ order on the other side.

\paragraph{Magnetic excitation spectrum.}

Each zigzag domain has a rectangular magnetic Brillouin zone. The triple-$\mathbf q$ state, by contrast, features a hexagonal Brillouin zone. The difference in the shape of the magnetic Brillouin zones leads to decisive consequence for the magnetic excitation spectrum.
To illustrate this, let us consider the hidden-SU(2)-symmetric point.
Figures~\ref{fig:spectrum-SU2}(b--g) compare the magnon spectrum in the zigzag state (top row) with those of the triple-$\mathbf q$ state (bottom row), for three different paths within the first crystallographic Brillouin zone, in linear spin-wave theory.
These paths cover points that are related by the symmetry of the triple-$\mathbf q$ state, but not of the zigzag state.
This allows the identification of triple-$\mathbf q$ order: While the triple-$\mathbf q$ magnon spectrum is symmetric with respect to these paths, the zigzag spectrum is not.
We emphasize that this symmetry argument is independent of a particular modeling, and applies equally to excitation beyond linear spin-wave theory, such as magnon bound states or lower bounds of magnetic excitation continua. It does not apply, however, to scattering intensities, which depend on absolute momenta.

Figure~\ref{fig:spectrum-NCTO} shows the low-temperature neutron scattering intensity of \NCTO\ along the same paths as in Fig.~\ref{fig:spectrum-SU2}(b--g), compiled from the experimental data initially published in Ref.~\cite{yao22a}. Remarkably, all visible magnon bands are perfectly symmetric along these trajectories, clearly indicating a hexagonal magnetic Brillouin zone with an $8$-site real-space unit cell.
Moreover, several qualitative features of the measured spectrum can be understood from the triple-$\mathbf q$ spectrum at the hidden-SU(2)-symmetric point. This includes the dispersive low-energy and flat high-energy bands, the number of excitation energies at the $\boldsymbol{\Gamma}$ point, as well as the small gap $\sim 1~\text{meV}$ at both $\boldsymbol{\Gamma}$ and $\mathbf M$. The latter can be considered as pseudo-Goldstone mode arising from the explicit breaking of the hidden-SU(2) symmetry, and be a measure of proximity to the hidden-SU(2)-symmetric point.

\begin{figure}[tb]
\includegraphics[width=\linewidth]{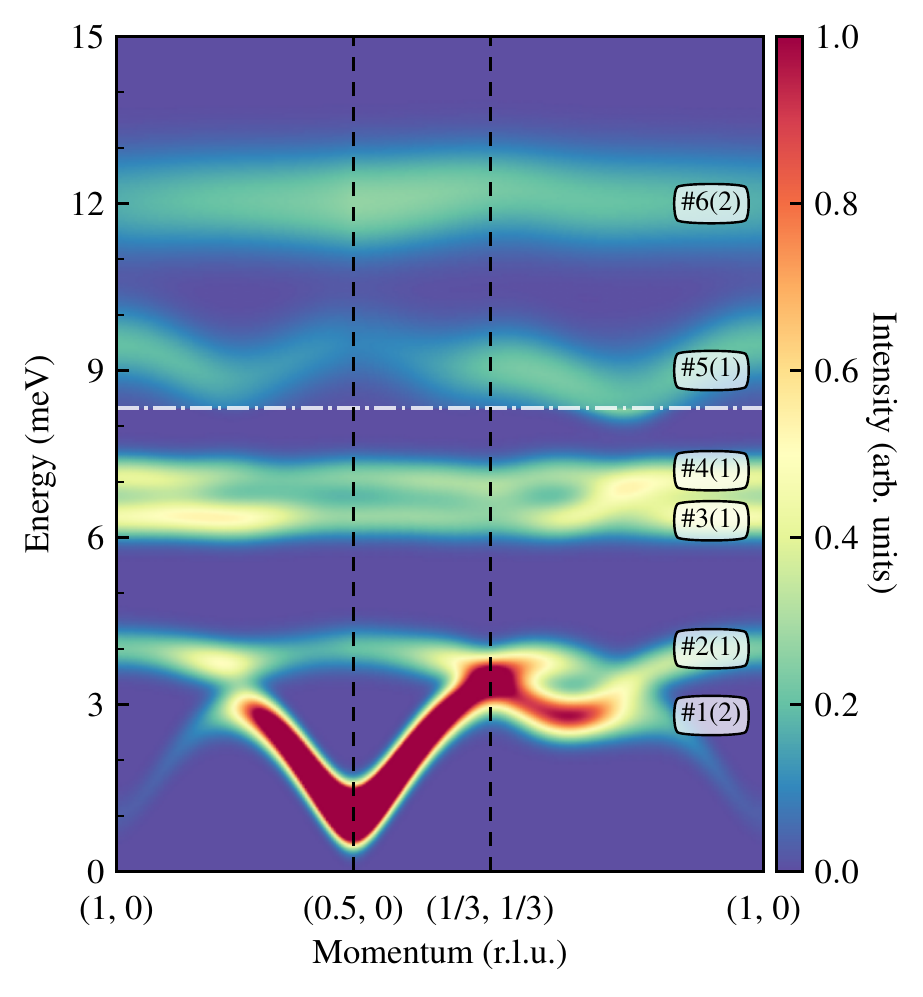}
\caption{Magnetic excitation spectrum from linear spin-wave theory of best-fit model, Eq.~\eqref{eq:fit}, to be compared with Fig.~3(a) of Ref.~\cite{yao22a}. The ground state is optimized to acquire the lowest energy, realizing the triple-$\mathbf q$ order shown in the lower inset of Fig.~\ref{fig:spectrum-SU2}(a). Different full width at half maxima (FWHM) are used in simulations at higher and lower energies for better comparison with experimental data: FWHM($E>8.3~\text{meV}$) = 1.18 meV, FWHM($E<8.3~\text{meV}$) = 0.59 meV. Branch indices are indicated after \cite{yao22a}, along with their practical degeneracy (defined as the number of branches within the FWHM).}
\label{fig:spectrum-Hmodel}
\end{figure}

\paragraph{Effective model for \NCTO.}

The above comparison suggests that the magnetic excitation spectrum in the low-temperature ordered phase of \NCTO\ can be modeled using a parameter set proximate to the hidden-SU(2)-symmetric point.
In linear-spin-wave theory, the ring exchange can be mapped to an effective local field along the local-moment axis, together with a renormalization of the bilinear couplings~\cite[(c)]{supplemental}.
In order to simplify the fitting algorithm, and to avoid the ambiguity arising from the various symmetry-allowed nonbilinear terms, we here constrain ourselves to an effective modeling using the local-field term directly, mimicking a ring exchange, i.e., $\mathcal H_\text{nbl} \simeq - h \sum_i \mathbf{n}_i \cdot \mathbf{S}_i$, with unit vector $\mathbf n_i$ along the triple-$\mathbf q$ spin direction at the hidden-SU(2)-symmetric point.
An alternative fitting procedure using the full ring exchange of Eq.~\eqref{eq:ring} leads to similar results~\cite[(d)]{supplemental}.
Optimizing the parameters with respect to the three spin-wave modes indicated by black lines in Fig.~\ref{fig:spectrum-NCTO} (see \cite[(d)]{supplemental} for details) leads to our best-fit model,
\begin{align} \label{eq:fit}
& (J_1, K_1, \Gamma_1, \Gamma'_1, J_2^A, J_2^B, J_3)_\text{fit} = 
    \nonumber \\ & \quad
(1.23, -8.29, 1.86, -2.27, 0.32, -0.24, 0.47)\,\text{meV}\,,
\end{align}
and $h = 0.88\,\text{meV}$, for $S=1/2$.
The resulting magnetic excitation spectrum is shown in Fig.~\ref{fig:spectrum-Hmodel}, to be compared with Fig.~3(a) of Ref.~\cite{yao22a}.
Importantly, the computed spectrum does not only reflect favorably the three fitting targets (modes \#1, \#3, and \#4), but moreover reproduces remarkably well also the other experimentally observed modes \#2, \#5, and \#6, both in terms of their bandwidths as well as their absolute energies.
Further comparisons between theoretical and experimental results, as well as a juxtaposition with results of a best-fit model with zigzag ground state, are presented in~\cite[(e)]{supplemental}.

\paragraph{Discussion.}

Our best-fit model suggests that the most dominant bilinear perturbation away from the hidden-SU(2)-symmetric point is towards the ferromagnet Kitaev limit. 
In~\cite[(f)]{supplemental}, we present extensive exact diagonalization calculations that show that \NCTO\ is in fact similarly close to the Kitaev quantum spin liquid regime as other intensely-studied Kitaev candidate materials, including Na$_2$IrO$_3$ and $\alpha$-RuCl$_3$.
Hence, while the low-temperature behavior in the triple-$\mathbf q$ phase can be well understood within the picture of approximate hidden-SU(2) symmetry, the nearby Kitaev quantum spin liquid should be expected to play an essential role at intermediate temperature and in external magnetic fields, when the magnetic order is melted~\cite{yao20, lin21, chen21, lee21, hong21,  yang22}.
In fact, the effective energy scale set by the Kitaev coupling of our best-fit model $|K_1|/k_\mathrm{B} \simeq 8.3\,\text{meV}/k_\mathrm{B} \sim 100\,\text{K}$ is well above the N\'eel temperature $T_\mathrm{N} \simeq 30\,\text{K}$~\cite{bera17, chen21, yao22a}, leaving an intermediate temperature regime that might be best described as a Kitaev paramagnet, similar to what has been observed in $\alpha$-RuCl$_3$~\cite{nasu16, banerjee16, banerjee17, do17}.
Furthermore, the noncollinear zero-field ordering implies increased magnetic frustration in external fields, as not all spins can cant homogeneously towards the field axis~\cite{janssen17, balz21}. 
As a consequence, a spin model proximate to the hidden-SU(2)-symmetric point generically features a metamagnetic transition between the triple-$\mathbf q$ state at small fields and a canted zigzag state at intermediate fields, before the transition towards the paramagnetic state at high fields, see~\cite[(g)]{supplemental} for details.
A recent variational Monte Carlo study also suggests a possible field-induced transition from a triple-$\mathbf q$ state at small fields towards a field-induced quantum spin liquid at intermediate in-plane fields~\cite{wang23}.
We expect these features of our minimal models to explain at least some of the various thermal and field-induced transitions observed in \NCTO~\cite{chen21, lee21, hong21, hong23, kocsis23}. 

\paragraph{Conclusions.}

We have demonstrated how a multi-$\mathbf q$ state that features Bragg peaks at symmetry-related points in the Brillouin zone can be distinguished experimentally from the corresponding multi-domain single-$\mathbf q$ state.
While the Bragg-peak pattern of both states can be identical, the difference in shape of the magnetic Brillouin zones leads to different symmetries of the magnetic excitation spectra.
Applying this approach to the honeycomb magnet \NCTO\ allows the identification of the material's low-temperature magnetic order: a triple-$\mathbf q$ state with a hexagonal $8$-site magnetic unit cell and a vortex spin structure around 1/4 of the elemental plaquettes.
%
%
Our result reveals \NCTO\ as an example of a Kitaev magnet proximate to a hidden-SU(2)-symmetric point. This explains not only the small band gap of the pseudo-Goldstone mode, but also the dispersive low-energy and flat high-energy bands observed in the spectrum.
The most dominant perturbation away from the hidden-SU(2)-symmetric point, however, is towards the ferromagnet Kitaev limit, and plays an essential role at intermediate temperature and in external magnetic fields.
This represents an intriguing question for future research.
In a broader context, our results suggest that the symmetry of the magnetic excitation spectrum may be worth reanalyzing also in other Kitaev magnets, such as \NCSO~\cite{li22}, or again $\alpha$-RuCl$_3$~\cite{johnson15, sears15, sears17, banerjee17, sears20, balz21}. 
The latter displays multiple uniform Bragg peaks in fields below 1--2~T~\cite{sears17, banerjee18}, a fact that was previously interpreted in terms of domain mixing arising from bond anisotropies~\cite{wolter17, lampenkelley18}, but could in principle also stem from multi-$\mathbf q$ order.
%

\begin{acknowledgments}
\paragraph{Acknowledgments.}
%
We thank
Fakher Assaad,
Bernd B\"uchner,
Jeroen van den Brink,
Matthias Gillig,
Christian Hess,
Xiaochen Hong,
Vladik Kataev,
Vilmos Kocsis,
Zheng-Xin Liu,
Satoshi Nishimoto,
Matthias Voj\-ta,
Jiucai Wang,
Christoph Wellm, and
Anja Wolter
for illuminating discussions and collaboration on related work.
We are grateful to Weiliang Yao for assistance on extracting some of the neutron scattering data originally published in \cite{yao22a}.
The data were obtained with the support of Kazuki Iida and Kazuya Kamazawa at the MLF, J-PARC, Japan, under a user program (Proposal No.~2019B0062).
Spin-wave calculations involved in the fitting procedure were performed on a computing cluster at Peking University using SpinW~\cite{toth15}.
Exact diagonalization calculations were performed on a computing cluster at TU Dresden using QuSpin~\cite{weinberg17, weinberg19}.
The work of WGFK and LJ is supported by the Deutsche Forschungsgemeinschaft (DFG) through SFB 1143 (A07, Project No.~247310070), the W\"{u}rzburg-Dresden Cluster of Excellence {\it ct.qmat} (EXC 2147, Project No.~390858490), and the Emmy Noether program (JA2306/4-1, Project No.~411750675).
The work of WC, XJ and YL is partly supported by the National Natural Science Foundation of China (NSFC, Grant No.\ 12061131004).

\paragraph{Note added.}
%
A parallel work reports in-field neutron diffraction data of \NCTO, which are consistent with our results~\cite{yao23}.

\end{acknowledgments}

\bibliographystyle{longapsrev4-2}
\bibliography{ncto-duality}


\end{document}


\title{Supplemental Material for ``Triple-q Order in Na\textsubscript{2}Co\textsubscript{2}TeO\textsubscript{6} from Proximity to Hidden-SU(2)-Symmetric Point''}

\author{Wilhelm G.\ F.\ Kr\"uger}
\thanks{WGFK and WC contributed equally to this work.}
\affiliation{Institut f\"ur Theoretische Physik and W\"urzburg-Dresden Cluster of Excellence ct.qmat, TU Dresden, 01062 Dresden, Germany}

\author{Wenjie Chen}
\thanks{WGFK and WC contributed equally to this work.}
\affiliation{International Center for Quantum Materials, School of Physics, Peking University, Beijing 100871, China}

\author{Xianghong Jin}
\affiliation{International Center for Quantum Materials, School of Physics, Peking University, Beijing 100871, China}

\author{Yuan Li}
\affiliation{International Center for Quantum Materials, School of Physics, Peking University, Beijing 100871, China}
\affiliation{Collaborative Innovation Center of Quantum Matter, Beijing 100871, China}

\author{Lukas Janssen}
\affiliation{Institut f\"ur Theoretische Physik and W\"urzburg-Dresden Cluster of Excellence ct.qmat, TU Dresden, 01062 Dresden, Germany}

\begin{abstract}
The Supplemental Material contains 
details about the $\mathcal T_1 \mathcal T_4$ duality transformation,
a discussion of perturbations away from the hidden-SU(2)-symmetric point, 
the spin-wave expansion of higher-order exchange terms,
details of the fitting algorithm to optimize the magnetic exchange couplings,
further comparisons between calculated and measured magnetic excitations,
a discussion of the proximity of our best-fit model for \NCTO\ to the Kitaev quantum spin liquid,
and the demonstration of metamagnetic transitions between triple-$\mathbf q$ and canted zigzag states in external fields.
\end{abstract}

\date{\today}

\maketitle

\section{\({\boldsymbol{\mathcal T}}\)\textsubscript{\!1}\({\boldsymbol{\mathcal T}}\)\textsubscript{\!4} duality transformation}

In this supplemental section, we discuss the duality transformation and the single-$\mathbf q$ and triple-$\mathbf q$ orders that are dual to the N\'eel antiferromagnets with staggered magnetizations in [001] and [111] directions, respectively.

\begin{figure}[b!]
\includegraphics[scale=1]{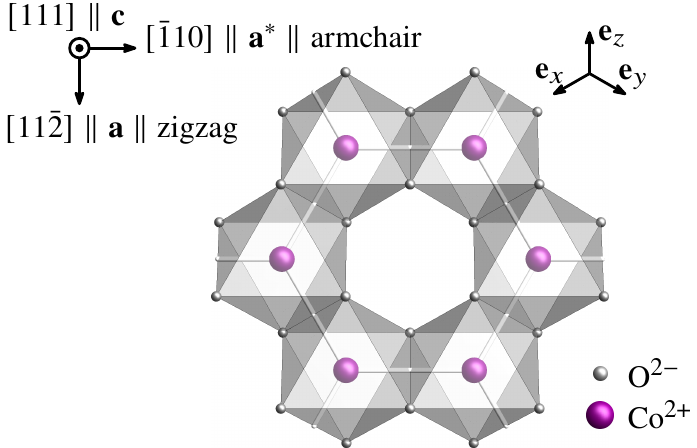}
\caption{Co$^{2+}$ honeycomb layers in \NCTO. We assume a crystal structure with undistorted honeycomb planes and in-plane crystallographic axes $\mathbf a$ and $\mathbf b$, and out-of-plane axis $\mathbf c \perp \mathbf a, \mathbf b$. The cubic spin-space basis vectors are denoted as $\mathbf e_x$, $\mathbf e_y$, and $\mathbf e_z$. They are oriented along Co-O bonds, such that $\mathbf c \propto \mathbf e_x + \mathbf e_y +\mathbf e_z \parallel  [111]$.}
\label{fig:axes}
\end{figure}

\begin{figure}[b!]
\includegraphics[scale=1]{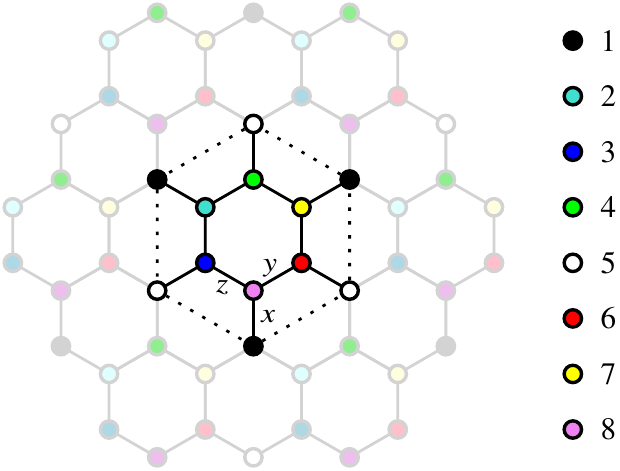}
\caption{8-site supercell (dotted hexagon) and sublattices (colored dots) of $\mathcal T_1 \mathcal T_4$ duality transformation. The transformation maps the N\'eel antiferromagnet with $\mathbf S_i \propto \pm \mathbf e_z$ to the $z$-zigzag state and the N\'eel antiferromagnet with $\mathbf S_i  \propto \pm (\mathbf e_x + \mathbf e_ y + \mathbf e_z)$ to the triple-$\mathbf q$ state.}
\label{fig:duality}
\end{figure}

The duality transformation that connects the HK$\Gamma\Gamma'$ model for the original spins, with parameters as in Eq.~(\eqSUtwo) of the main text, to a pure Heisenberg model for the dual spins is given by a combination of two different duality transformations discussed in Ref.~\cite{chaloupka15}.
The first one is the well-known four-sublattice transformation $\mathcal T_4$~\cite{chaloupka10}. It corresponds to a site-dependent $\pi$ rotation around the cubic axes $[100]$, $[010]$, and $[001]$ of the spins on three of the four sublattices, with the spins on the fourth sublattice remaining invariant. 
Here, the cubic spin-space axes are defined as depicted in Fig.~\ref{fig:axes}.
The ensuing transformation $\mathcal T_1$ corresponds to a global $\pi$ rotation around the $[111]$ axis.
Under the combined $\mathcal T_1 \mathcal T_4$ transformation, a N\'eel antiferromagnet with staggered magnetization $\mathbf n \parallel [001]$ maps to a $z$-zigzag state with spin directions
\begin{align}
\mathbf S_i = 
\frac{S}{3}
\begin{cases}
+(2 \mathbf e_x + 2 \mathbf e_y - \mathbf e_z) & \text{for } i \in \text{sublattices } 1,6,7,8, \\
-(2 \mathbf e_x + 2 \mathbf e_y - \mathbf e_z) & \text{for } i \in \text{sublattices } 2,3,4,5,
\end{cases}
\end{align}
on the eight different sublattices indicated in Fig.~\ref{fig:duality}. 
The $z$-zigzag state features collinear order with antiferromagnetic $z$ bonds and ferromagnetic $x$ and $y$ bonds. It has a single Bragg peak at the center $\mathbf M_z$ of one of the edges of the first Brillouin zone.
In general, however, the duality transformation does not map the N\'eel state to a collinear single-$\mathbf q$ state.
For a staggered magnetization in the out-of-plane direction $\mathbf n \parallel [111]$, for instance, the N\'eel state is mapped to a triple-$\mathbf q$ state with spin directions
\begin{align} \label{eq:triple-Q}
\mathbf S_i = 
\frac{S}{3\sqrt{3}}
\begin{cases}
+3(\mathbf e_x + \mathbf e_y + \mathbf e_z) & \text{for } i \in \text{sublattice } 1, \\
-(\mathbf e_x + \mathbf e_y - 5 \mathbf e_z) & \text{for } i \in \text{sublattice } 2, \\
+(\mathbf e_x - 5 \mathbf e_y + \mathbf e_z) & \text{for } i \in \text{sublattice } 3, \\
+(-5 \mathbf e_x + \mathbf e_y + \mathbf e_z) & \text{for } i \in \text{sublattice } 4, \\
-3(\mathbf e_x + \mathbf e_y + \mathbf e_z) & \text{for } i \in \text{sublattice } 5, \\
+(\mathbf e_x + \mathbf e_y - 5 \mathbf e_z) & \text{for } i \in \text{sublattice } 6, \\
-(\mathbf e_x - 5 \mathbf e_y + \mathbf e_z) & \text{for } i \in \text{sublattice } 7, \\
-(-5 \mathbf e_x + \mathbf e_y + \mathbf e_z) & \text{for } i \in \text{sublattice } 8,\\
\end{cases}
\end{align}
on the eight sublattices indicated in Fig.~\ref{fig:duality}.
This triple-$\mathbf q$ state features a hexagonal eight-site magnetic unit cell with the two spins at its corners pointing along the out-of-plane direction $\pm [111]$, and the six spins around the elemental plaquette forming a vortex configuration, see lower inset of Fig.~\figspectrumSUtwo(a) in the main text.
This state was previously found as a ground state in a region of the phase diagram of the Heisenberg-Kitaev model in an external $[111]$ field in Ref.~\cite{janssen16}, where it was denoted as ``AF star''%
~\footnote{Note that the spin-space basis used in the plots of the spin configurations in Ref.~\cite{janssen16} is rotated by $\pi/2$ about the out-of-plane axis in comparison to the conventions employed here.}.

\section{Perturbations away from hidden-SU(2) point}

In this supplemental section, we discuss which kind of perturbations away from the hidden-SU(2)-symmetric point can lift the degeneracy between single-$\mathbf q$ and triple-$\mathbf q$ states in the SU(2) manifold that is dual to the ground state of the pure Heisenberg antiferromagnet.

\subsection{Bilinear perturbations}

\paragraph*{SU(2) degeneracy in classical limit.} 
We start by arguing that the SU(2) ground-state degeneracy that is exact at the hidden-SU(2)-symmetric point is not lifted by any bilinear interactions consistent with the lattice symmetries in the classical limit.
Technically, this property can be understood as a consequence of the fact that the $\mathcal T_4$ transformation effectively only changes signs of the various interaction terms on half of the bonds~\cite{sizyuk16}, while $\mathcal T_1$ corresponds to just a global spin rotation.
For concreteness, consider the first-neighbor $\Gamma'_1$ interaction on the six bonds $\langle 23 \rangle_x$, $\langle 24 \rangle_y$, $\langle 12 \rangle_z$, $\langle 18 \rangle_x$, $\langle 35 \rangle_y$, and $\langle 38 \rangle_z$  in the lower left half of the hexagonal eight-site supercell, Fig.~\ref{fig:duality}.
The SU(2) manifold that includes both zigzag and triple-$\mathbf q$ states can be parameterized by the components $(n_x, n_y, n_z)$ of the staggered-magnetization axis of the corresponding dual N\'eel antiferromagnet, with $n_x^2+n_y^2+n_z^2 = 1$.
The $\Gamma'_1$ interaction for states of the SU(2) manifold on the specified six bonds can then be written as
\begin{align}
%
%
\langle 23 \rangle_x &:&
S^x_2 S^z_3 
+ S^x_2 S^y_3 
+ (2\leftrightarrow3)
& = \frac{4S}{9} [2 n_x^2 + (n_y - n_z)^2]\,, \displaybreak[1] \nonumber\\
%
%
\langle 24 \rangle_y  &:&
S^y_2 S^z_4 
+ S^y_2 S^x_4 
+ (2\leftrightarrow4)
& = \frac{4S}{9} [2 n_y^2 + (n_z - n_x)^2]\,, \displaybreak[1] \nonumber\\
%
%
\langle 38 \rangle_z  &:& 
S^z_3 S^x_8 
+ S^z_3 S^y_8 
+ (3\leftrightarrow8)
& = \frac{4S}{9} [2 n_z^2 + (n_x - n_y)^2]\,, \displaybreak[1] \nonumber\\
%
%
\langle 18 \rangle_x  &:&
S^x_1 S^y_8 
+ S^x_1 S^z_8 
+ (1\leftrightarrow8)
& = \frac{4S}{9} [2 n_x^2 + (n_y + n_z)^2]\,, \displaybreak[1] \nonumber\\
%
%
\langle 35 \rangle_y  &:&
S^y_3 S^z_5 
+ S^y_3 S^x_5 
+(3\leftrightarrow5)
& = \frac{4S}{9} [2 n_y^2 + (n_z + n_x)^2]\,, \displaybreak[1] \nonumber\\
%
%
\langle 12 \rangle_z  &:&
S^z_1 S^x_2 
+ S^z_1 S^y_2 
+ (1\leftrightarrow2)
& = \frac{4S}{9} [2 n_z^2 + (n_x + n_y)^2]\,.
\end{align}
In the above, the first three lines correspond to the bonds along the central elemental plaquette in Fig.~\ref{fig:duality}, while the last three lines correspond to the bonds connecting a site on the central plaquette with a corner of the 8-site supercell.
Importantly, the signs of the mixed terms on the former set of bonds are opposite as compared to those on the latter, such that the sum of the $\Gamma'$ interaction on the six bonds yields $\frac{32}{9} (n_x^2 + n_y^2 + n_z^2)$, independent of the direction of the staggered-magnetization axis $\mathbf n$, featuring full SU(2) symmetry.
By inversion symmetry, the same cancellation of the SU(2)-breaking mixed terms occurs on the other six bonds within the 8-site unit cell, implying that the $\Gamma'$ interaction does not lift the SU(2) degeneracy of the states dual to the N\'eel state.
One can similarly show that the same type of cancellation occurs also for the $K_1$ and $\Gamma_1$ interactions.
In fact, in the classical limit, the above argument is not restricted to first-neighbor bonds, and we expect the degeneracy to hold for arbitrary long-range interactions consistent with the symmetries of the model.
In particular, we have explicitly verified that second-neighbor Dzyaloshinskii-Moriya interactions do not lift the degeneracy.

\paragraph*{Rotational-symmetry-breaking perturbations.}
By contrast, perturbations that break the $C_3^*$ symmetry of $2\pi/3$ spin-lattice rotation about the out-of-plane axis through a lattice site can lift the SU(2) degeneracy on the classical level.
These are expected to favor states for which the dual spins point along or perpendicular to one of the cubic axes, leading to single-$\mathbf q$ zigzag order.
We have verified this expectation for a number of different such symmetry-breaking terms explicitly.

\paragraph*{Sublattice-symmetry-breaking perturbations.}
As mentioned in the main text, there are two crystallographically inequivalent Co$^{2+}$ sites in \NCTO, thereby breaking the sublattice symmetry and leading to a small ferrimagnetic moment~\cite{yao20}. When the sublattice symmetry is broken, second-neighbor interactions can be different on the different next-neighbor bonds.
We have explicitly verified that such symmetry-breaking interactions do not lift the SU(2) degeneracy for bilinear interactions in the classical limit.

In sum, this suggests that on the level of bilinear exchange interactions, there are no perturbations that favor, in the classical limit, the triple-$\mathbf q$ states over the single-$\mathbf q$ states present in the SU(2) manifold.
In fact, except for rotational-symmetry-breaking perturbations, the classical SU(2) degeneracy that is symmetry required at the hidden-SU(2)-symmetric point remains intact upon inclusion of all discussed bilinear perturbations.
We remark that this does not necessarily imply that the ground state features an SU(2) degeneracy in the classical limit, as the SU(2) manifold can be shifted to higher energies in the presence of bilinear perturbations.

\paragraph*{Order from disorder.}
Thermal and quantum fluctuations, by contrast, do lift the degeneracy of the SU(2) manifold, when perturbations away from the hidden-SU(2)-symmetric point are taken into account.
It is widely expected from various explicit calculations in Heisenberg-Kitaev and other frustrated magnetic systems  that both thermal and quantum fluctuations favor collinear states over noncollinear states, as a consequence of the energetically steeper parameter space near noncollinear spin configurations~\cite{chaloupka10, rau18, consoli20}.
We have explicitly verified this to be also the case of the HK$\Gamma\Gamma'$ model at the hidden-SU(2)-symmetric point, Eq.~(\eqSUtwo) of the main text, upon the inclusion of first-neighbor Kitaev exchange perturbations.
This suggests that triple-$\mathbf q$ order does not arise from an order-from-disorder effect, but, if present, should be stabilized already on the classical level.

\subsection{On-site perturbations}
%
A perturbation that does lift the SU(2) degeneracy already on the classical level is an on-site magnetic field. 
In fact, an external field along the $[111]$ direction indeed favors the triple-$\mathbf q$ order over the single-$\mathbf q$ order. This was initially observed for the original Heisenberg-Kitaev model in Ref.~\cite{janssen16}, and we have explicitly verified that the same happens also in the present model with additional $\Gamma_1$ and $\Gamma'_1$ interactions.
The reason for this effect is that the single-$\mathbf q$ state present in the SU(2) manifold have sizable out-of-plane components and can therefore only insufficiently cant towards the out-of-plane axis. This is in contrast to the triple-$\mathbf q$ state, for which six out of eight spins in the supercell have small out-of-plane components in the zero-field configuration.
On the other hand, as originating from spin canting, the stabilization of the triple-$\mathbf q$ state is a higher-order effect, implying that the energy difference between canted zigzag and canted triple-$\mathbf q$ states is small.

A staggered ``internal'' field $\mathbf h_\text{stagg}$ along the out-of-plane direction, which can be thought of as modeling the sublattice symmetry breaking in \NCTO\ \cite{yao20},
\begin{align}
	\mathcal H_\text{stagg} = - \mathbf h_\text{stagg} \cdot \left[\sum_{i \in A} \mathbf S_i - \sum_{j \in B} \mathbf S_j\right],
\end{align}
where $A$ and $B$ denote the two sublattices of the honeycomb lattice, can also favor the triple-$\mathbf q$ state over the single-$\mathbf q$ states. %
However, similar to the case of the global external field, the energy gain arises from different spin canting processes of the different states involved, and as such is a higher-order effect in the field.
Indeed, for small fields $h_\text{stagg} = |\mathbf h_\text{stagg}| \ll A$ in the vicinity of the hidden-SU(2)-symmetric point, the energy gain per site is quadratic in the field strength,
\begin{align}
	\frac{E_\text{triple-$\mathbf q$}}{N A S^2} = -\frac{3}{2} - \frac{1}{12} \left(\frac{h_\text{stagg}}{A S}\right)^2 + \mathcal O(h_\text{stagg}^3)\,.
\end{align}

This implies that the canted triple- and single-$\mathbf q$ states will remain very close in energy unless $h_\text{stagg}$ becomes large.
We conclude that while in principle a staggered magnetization can account for a triple-$\mathbf q$ order, it will only do so if the material is very close to the hidden-SU(2)-symmetric point.

\subsection{Ring perturbations}

\begin{figure}[tb]
\includegraphics[width=\linewidth]{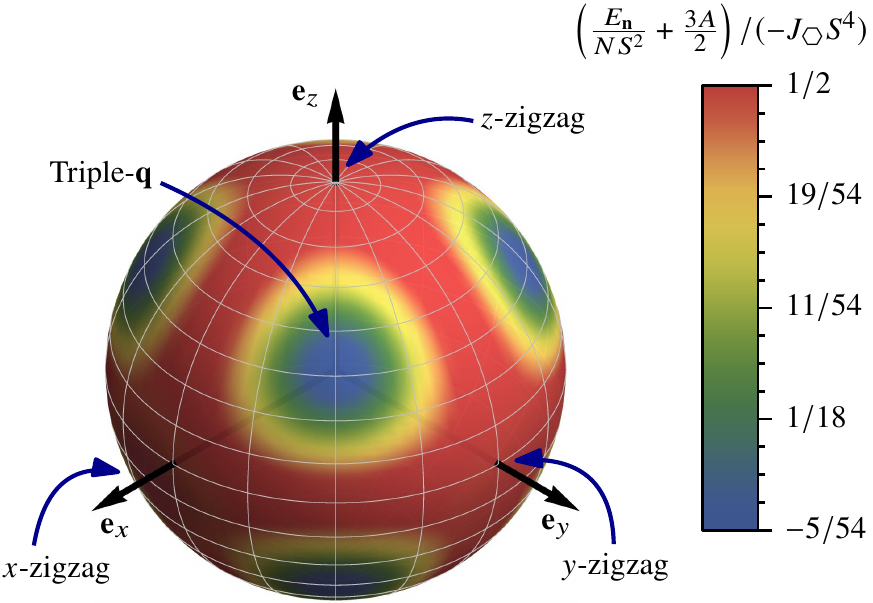}
\caption{Classical energy of states in the SU(2) manifold in presence of small ring exchange perturbation $\mathcal H_\text{nbl}$, Eq.~(\eqring) of the main text, parameterized by the spherical angles $\theta$ and $\varphi$ of the staggered-magnetization axis $\mathbf n = (\sin\theta \cos\varphi, \sin\theta\sin\varphi, \cos\theta)$ of the dual N\'eel state. For $J_{\hexagon}<0$, the minimal energy is obtained at $\mathbf n \parallel [111]$ or symmetry-related, corresponding to the triple-$\mathbf q$ state.}
\label{fig:energy}
\end{figure}

As discussed in the main text, exchange interactions beyond on-site and bilinear terms are important in a number of $3d$ materials, including various chromium-, manganese-, and copper-based magnets~\cite{kvashnin20,fedorova15,dallapiazza12,larsen19}.
This applies in particular to ring exchange perturbations, which arise from higher-order corrections in the strong-coupling expansion of the Hubbard model~\cite{takahashi77, macdonald88}.
As the ring exchange couples more that just two spins, it lifts the classical SU(2) degeneracy of the states that are dual to the N\'eel state already on the linear level $\propto J_{\hexagon}$. 
In fact, we find the classical energy $E_{\mathbf n}$ of the state that is dual to the N\'eel state with staggered-magnetization axis $\mathbf n = (\sin\theta \cos\varphi, \sin\theta\sin\varphi, \cos\theta)$ in the presence of a small ring exchange perturbation as 
\begin{align}
\frac{E_{\mathbf n}}{N A S^2} & = -\frac{3}{2} - \frac{J_{\hexagon} S^4}{16 A}
\bigl[32 \sin ^4\theta \cos ^2\theta  \cos 4 \varphi + \cos 2 \theta 
\nonumber \\ & \quad
+ 2 \cos 4 \theta - \cos 6 \theta + 6 \bigr] + \mathcal O(J_{\hexagon}^2)\,,
\end{align}
which evaluates to $E_{[111]}/(N A S^2) = -3/2 + 5 J_{\hexagon} S^4/(54 A)$ for the triple-$\mathbf q$ state and $E_{[001]}/(N A S^2) = -3/2 - J_{\hexagon} S^4/(2A)$ for the single-$\mathbf q$ zigzag states.
The energy as function of $\theta$ and $\varphi$ is depicted in Fig.~\ref{fig:energy}. In agreement with the phase diagram shown in Fig.~\figspectrumSUtwo(a) in the main text, for $J_{\hexagon}<0$, the minimal energy is obtained at $\mathbf n \parallel [111]$ or symmetry-related, corresponding to the triple-$\mathbf q$ state, while the zigzag state is favored for $J_{\hexagon} > 0$.

%

\section{Spin-wave theory of higher-order exchange}

In this supplemental section, we describe the spin-wave expansion of generic exchange terms beyond the bilinear order, such as ring exchanges, and demonstrate that such terms can be effectively modeled, on the level of linear spin-wave theory, with on-site and renormalized bilinear exchange interactions.

We employ the Holstein-Primakoff representation of the spin operator $\mathbf{S}_i$ on lattice site $i$ with magnon creation and annihilation operators $\xi_i^\dagger$ and $\xi_i$,
\begin{equation}
  S^-_i=\xi^\dagger_i\sqrt{2S-\xi^\dagger_i\xi^\pd_i},\quad
  S^+_i=\sqrt{2S-\xi^\dagger_i\xi^\pd_i}\xi^\pd_i,\quad
  S^3_i=S-\xi^\dagger_i\xi^\pd_i.
\label{2:formel:HolsteinPrimakofftransformation}
\end{equation}
Here, $S^\pm_i = S^1_i \pm \rmi S^2_i$ are spin ladder operators, and $S^\alpha_i$, $\alpha = 1,2,3$, denote the components of the spin operator in the local basis given by the classical spin direction, such that $\langle S^x_i \rangle = \langle S^y_i \rangle = 0$ in the semiclassical limit.
The magnon operators fulfill the fundamental commutation relation $[\xi^\pd_i, \xi^\dagger_j ] = \delta_{ij}$.
In the large-$S$ limit, the spin operators are simplified to
\begin{equation}
  \begin{split}
  S^-_i=\sqrt{2S}\xi^\dagger_i,\quad
  S^+_i=\sqrt{2S}\xi^\pd_i,\quad
  S^3_i=S-\xi^\dagger_i\xi^\pd_i.
  \end{split}
\end{equation}
Let us consider a spin-exchange term $\mathcal H_{i_1\dots i_n}$ involving $n$ magnetic sites $i_1,  \dots, i_n$.
On-site terms correspond to $n=1$, bilinear exchange terms correspond to $n=2$, and ring exchange terms on elemental plaquettes of the honeycomb lattice correspond to $n=6$.
The $1/S$ expansion of such generic spin-exchange term reads
\begin{align}
  \mathcal H_{i_1 \dots i_n} & = 
  \mathcal H_{i_1 \dots i_n} \bigr|_{0} 
  + \partial \mathcal H_{i_1 \dots i_n} \bigr|_{0} S^{-1/2}
  \nonumber \\ & \quad
  + \frac{1}{2} \partial^2 \mathcal H_{i_1 \dots i_n} \bigr|_{0} S^{-1}
  + \dots\,,
\end{align}
where we have abbreviated $\partial \equiv \partial/[\partial(S^{-1/2})]$ and $(\,\cdots\,)\bigr|_0 \equiv (\,\cdots\,)\bigr|_{S^{-1/2} \to 0}$.
In the above equation, $\mathcal H_{i_1 \dots i_n} \bigr|_{0} =  E_\text{cl} \propto S^n$ contains the classical energy, 
$\partial \mathcal H_{i_1 \dots i_n} \bigr|_{0}$ contains terms linear in the magnon operators $\xi_i$ and $\xi_i^\dagger$, which cancel when expanding around a local minimum of the classical energy, 
and $\partial^2 \mathcal H_{i_1 \dots i_n} \bigr|_{0}$ contains quadratic terms in $\xi_i$ and $\xi_i^\dagger$.
The latter determine the magnon spectrum in linear spin-wave theory.
Let us discuss these in more detail.
An on-site term $\mathcal H_i = \sum_{\alpha} H_{i}^{\alpha}S_i^\alpha$ at site~$i$, such as a local field term, can be written as
\begin{equation} \label{eq:SWT-onsite}
  \partial^2 \mathcal H_{i} = \sum_{\alpha} H_{i}^{\alpha}\partial^2{S}_i^\alpha\,.
\end{equation}
In the equation above and in what follows, the semiclassical limit $S^{-1/2} \to 0$ is implicitly understood, i.e., $\partial^2 \mathcal H_{i} \to \partial^2 \mathcal H_{i}\bigr|_0$.
A bilinear term $\mathcal H_{ij} = \sum_{\alpha\beta}H_{ij}^{\alpha\beta}S_i^\alpha S_j^\beta $ on a bond connecting sites $i$ and $j$ can be written as
\begin{align} \label{eq:SWT-bilinear}
  \partial^2 \mathcal H_{ij}
  = \sum_{\alpha\beta}H_{ij}^{\alpha\beta} \partial^2 (S^\alpha_i S^\beta_j)\,.
\end{align}
Finally, a sixth-order term $\mathcal H_{ijklmn} = \sum_{\alpha\beta\gamma\delta\epsilon\zeta}H_{ijklmn}^{\alpha\beta\gamma\delta\epsilon\zeta}S_i^\alpha S_j^\beta S_k^\gamma S_l^\delta S_m^\epsilon S_n^\zeta$, 
such as the honeycomb-lattice ring exchange term in Eq.~(\eqring) of the main text, can be written as
\begin{align} \label{eq:SWT-ring}
  \partial^2 \mathcal H_{ijklmn} & = 
  \sum_{\alpha\beta\gamma\delta\epsilon\zeta}H_{ijklmn}^{\alpha\beta\gamma\delta\epsilon\zeta}\Big\{
    S_k^\gamma S_l^\delta S_m^\epsilon S_n^\zeta \partial^2(S^\alpha_i S^\beta_j)
   \nonumber \displaybreak[1] \\ & \!\!
  + S_j^\beta S_l^\delta S_m^\epsilon S_n^\zeta \partial^2(S^\alpha_i S^\gamma_k)
  + S_j^\beta S_k^\gamma S_m^\epsilon S_n^\zeta \partial^2(S^\alpha_i S^\delta_l)
   \nonumber \displaybreak[1] \\ & \!\!
  + S_j^\beta S_k^\gamma S_l^\delta S_n^\zeta \partial^2(S^\alpha_i S^\epsilon_m)
  + S_j^\beta S_k^\gamma S_l^\delta S_m^\epsilon \partial^2(S^\alpha_i S^\zeta_n)
   \nonumber \displaybreak[1] \\ & \!\!
  + S_i^\alpha S_l^\delta S_m^\epsilon S_n^\zeta \partial^2(S^\beta_j S^\gamma_k)
  + S_i^\alpha S_k^\gamma S_m^\epsilon S_n^\zeta \partial^2(S^\beta_j S^\delta_l)
   \nonumber \displaybreak[1] \\ & \!\!
  + S_i^\alpha S_k^\gamma S_l^\delta S_n^\zeta \partial^2(S^\beta_j S^\epsilon_m)
  + S_i^\alpha S_k^\gamma S_l^\delta S_m^\epsilon \partial^2(S^\beta_j S^\zeta_n)
   \nonumber \displaybreak[1] \\ & \!\!
  + S_i^\alpha S_j^\beta S_m^\epsilon S_n^\zeta \partial^2(S^\gamma_k S^\delta_l)
  + S_i^\alpha S_j^\beta S_l^\delta S_n^\zeta \partial^2(S^\gamma_k S^\epsilon_m)
   \nonumber \displaybreak[1] \\ & \!\!
  + S_i^\alpha S_j^\beta S_l^\delta S_m^\epsilon \partial^2(S^\gamma_k S^\zeta_n)
  + S_i^\alpha S_j^\beta S_k^\gamma S_n^\zeta \partial^2(S^\delta_l S^\epsilon_m)
   \nonumber \displaybreak[1] \\ & \!\!
  + S_i^\alpha S_j^\beta S_k^\gamma S_m^\epsilon \partial^2(S^\delta_l S^\zeta_n)
  + S_i^\alpha S_j^\beta S_k^\gamma S_l^\delta \partial^2 (S^\epsilon_m S^\zeta_n)
   \nonumber \displaybreak[1] \\ & \!\!\!\!\!\!\!\!
  - 4\big[
  S_j^\beta S_k^\gamma S_l^\delta S_m^\epsilon S_n^\zeta (\partial^2{S}_i^\alpha)
  + S_i^\alpha S_k^\gamma S_l^\delta S_m^\epsilon S_n^\zeta (\partial^2{S}_j^\beta) 
   \nonumber \displaybreak[1] \\ & \!\!
  + S_i^\alpha S_j^\beta S_l^\delta S_m^\epsilon S_n^\zeta (\partial^2{S}_k^\gamma) 
  + S_i^\alpha S_j^\beta S_k^\gamma S_m^\epsilon S_n^\zeta (\partial^2{S}_l^\delta) 
   \nonumber \displaybreak[1] \\ & \!\!
  + S_i^\alpha S_j^\beta S_k^\gamma S_l^\delta S_n^\zeta (\partial^2{S}_m^\epsilon)
  + S_i^\alpha S_j^\beta S_k^\gamma S_l^\delta S_m^\epsilon (\partial^2{S}_n^\zeta)
   \big]
  \Big\}\,.
\end{align}
Consider the first 15 terms in the curly brackets $\{\,\cdots\,\}$ of the above equation.
In each term, the first four factors, such as $S_k^\gamma S_l^\delta S_m^\epsilon S_n^\zeta$ in the first term, become real numbers in the semiclassical limit, given by the classical spin configuration.
The fifth and last factors in each term, such as $\partial^2(S^\alpha_i S^\beta_j)$ in the first term, are operators quadratic in $\xi_i$ and $\xi_i^\dagger$. They are precisely of the form of Eq.~\eqref{eq:SWT-bilinear}.
These 15 terms can therefore be written as spin-wave expansions of bilinear terms $\partial^2 \mathcal H_{i'j'}$ with $i',j' \in \{i,j,k,l,m,n\}$ and effective couplings $H_{i'j'}^{\alpha \beta}$. For instance, for the first term with $i'=i$ and $j'=j$, we obtain an effective bilinear coupling
$H_{ij}^{\alpha \beta} = \sum_{\gamma\delta\epsilon\zeta} H_{ijklmn}^{\alpha\beta\gamma\delta\epsilon\zeta} S_k^\gamma S_l^\delta S_m^\epsilon S_n^\zeta$.
Consider now the 6 terms in the square bracket $[\,\cdots\,]$ in Eq.~\eqref{eq:SWT-ring}.
The last factors in these terms, such as $\partial^2(S^\alpha_i)$ in the first one, are precisely of the form of Eq.~\eqref{eq:SWT-onsite}.
These 6 terms can therefore be written as spin-wave expansions of on-site terms $\partial^2 \mathcal H_{i'}$ with $i' \in \{ i,j,k,l,m,n\}$ and effective local fields $H_{i'}^\alpha$.
For instance, for the first term with $i' = i$, we obtain an effective on-site term with coefficient
$H_{i}^\alpha = -4 S_j^\beta S_k^\gamma S_l^\delta S_m^\epsilon S_n^\zeta$.

In sum, on the level of linear spin-wave theory, an effective model with renormalized bilinear exchange couplings and additional on-site couplings exists for any given six-spin exchange term, such as the honeycomb-lattice ring exchange. 
We note that the couplings on the different bonds respect the symmetries of the magnetic lattice only, which will generically be less than those of the crystallographic lattice.
For instance, for the ring exchange term arising from the Hubbard-model expansion, Eq.~(\eqring) in the main text, with a triple-$\mathbf q$ ground state for small $J_{\hexagon} < 0$, the resulting effective on-site term arising from Eq.~\eqref{eq:SWT-ring} corresponds to a local field that is along the local spin axis, but opposite to the classical spin configuration.

Along the same lines, a general $N$-spin exchange term can mapped, on the level of linear spin-wave theory, to an effective model with $\binom{N}{2}=\frac{1}{2}N(N-1)$ bilinear terms and $N$ linear terms, the latter with a prefactor $-(N-2)$.
This mapping to an effective model allows to simulate any higher-order spin exchange using publicly available codes, such as SpinW~\cite{toth15}.

\section{Details of fitting algorithm}

In this supplementary section, we give details of the fitting algorithm to obtain an effective spin model that reproduces the measured magnetic excitation spectrum of \NCTO.
This has been achieved by global optimization using a genetic algorithm, followed by an ensuing local optimization.
For the best-fit model discussed in the main text, the model parameters include seven bilinear interactions $(J_1, K_1, \Gamma_1, \Gamma'_1, J_2^A, J_2^B, J_3)$, as well as one parameter $h$ for the local field $\mathbf h_i = h \mathbf n_i$, with  unit vector $\mathbf n_i$ along the triple-$\mathbf q$ spin direction at the hidden-SU(2)-symmetric point, see Eq.~\eqref{eq:triple-Q}. Hence, the dimension of parameter space is 8.

In the genetic algorithm, we set the population size of each generation to 320 individuals.
Each individual corresponds to an 8-component parameter set.
For each individual, the spin ordering is optimized to ensure that the spin-wave spectrum is computed with respect to the correct ground state.
The fitness value of each individual is calculated by comparing the spin-wave spectrum of the corresponding model with the fitting targets, as discussed below.
The initial population is randomly generated within the constrained 8-dimensional parameter space, and then evolves for 50 generations.
In each generation, individuals are randomly chosen to produce children for the next generation.
While mutations can maintain the genetic diversity, keeping the algorithm from falling into local minima, individuals with bad fitness values are generally washed out by natural selection.
The best-fit model from the last generation is then examined and put into refinement by a local optimization.
This fitting algorithm was applied multiple times independently to ensure the consistency of the best-fit model.
All acceptable models turned out to feature a triple-$\mathbf q$ ground state. In particular, it has not been possible to model the magnetic excitation spectrum using a spin-wave expansion around a single-$\mathbf q$ state, such as zigzag.

\begin{figure*}[tbp]
\includegraphics[width=\linewidth]{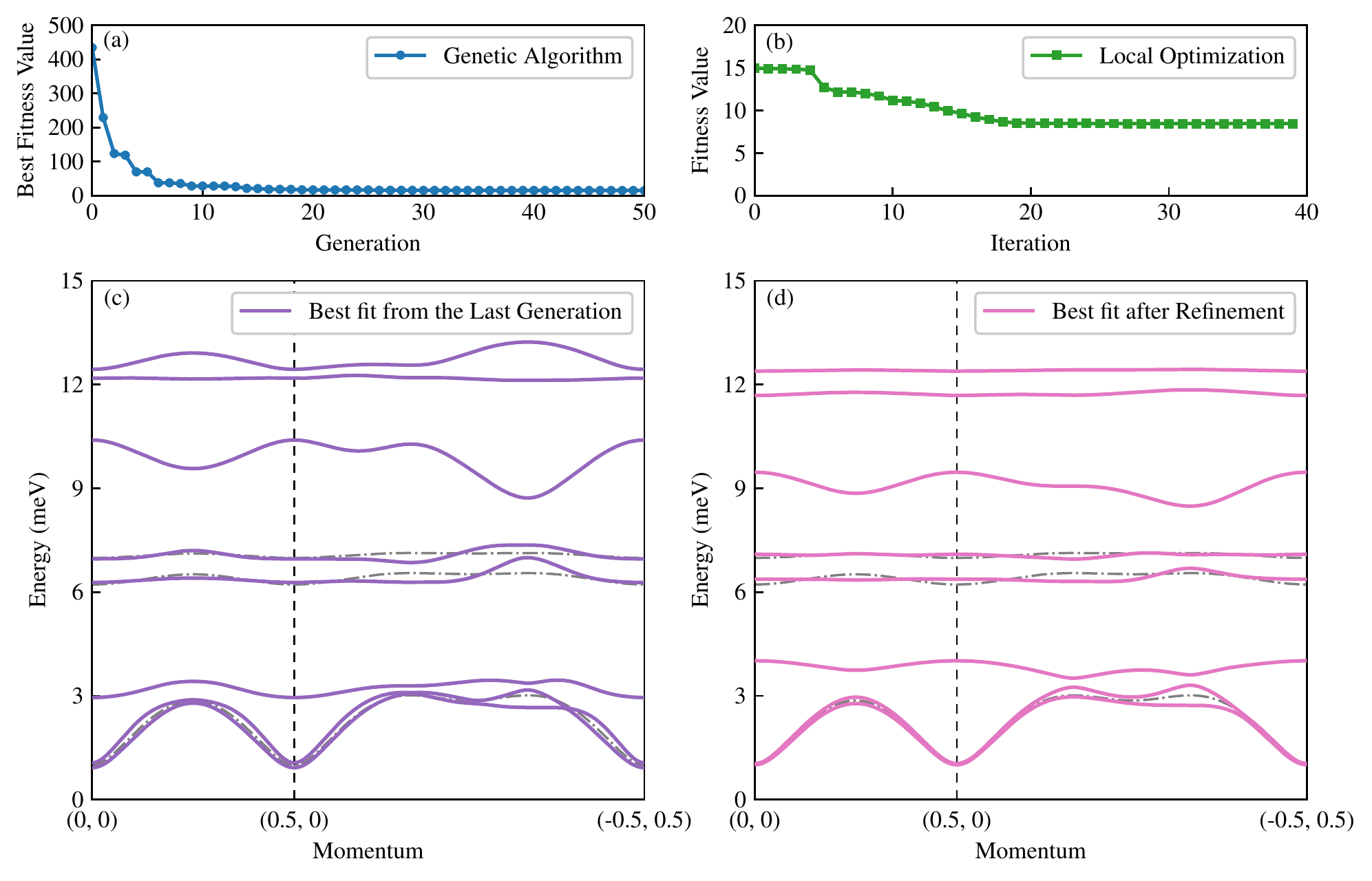}
\caption{Fitting process of the spin-wave model in the 8-dimensional parameter space. (a) Global optimization with the genetic algorithm. The best fitness value of one generation is defined as the smallest fitness value out of all individuals. (b) Local optimization starting with the best-fit model from the genetic algorithm. (c) Spin-wave dispersions of the best-fit model from the genetic algorithm. (d) Spin-wave dispersions of the best-fit model after additional local optimization, which is the triple-$\mathbf q$ model mentioned in the main text. Gray dash-dotted curves indicate the three spin-wave dispersions that served as fitting targets.}
\label{fig:GA_refine_comparison_supp}
\end{figure*}

\begin{figure*}[tbp]
\includegraphics[width=\linewidth]{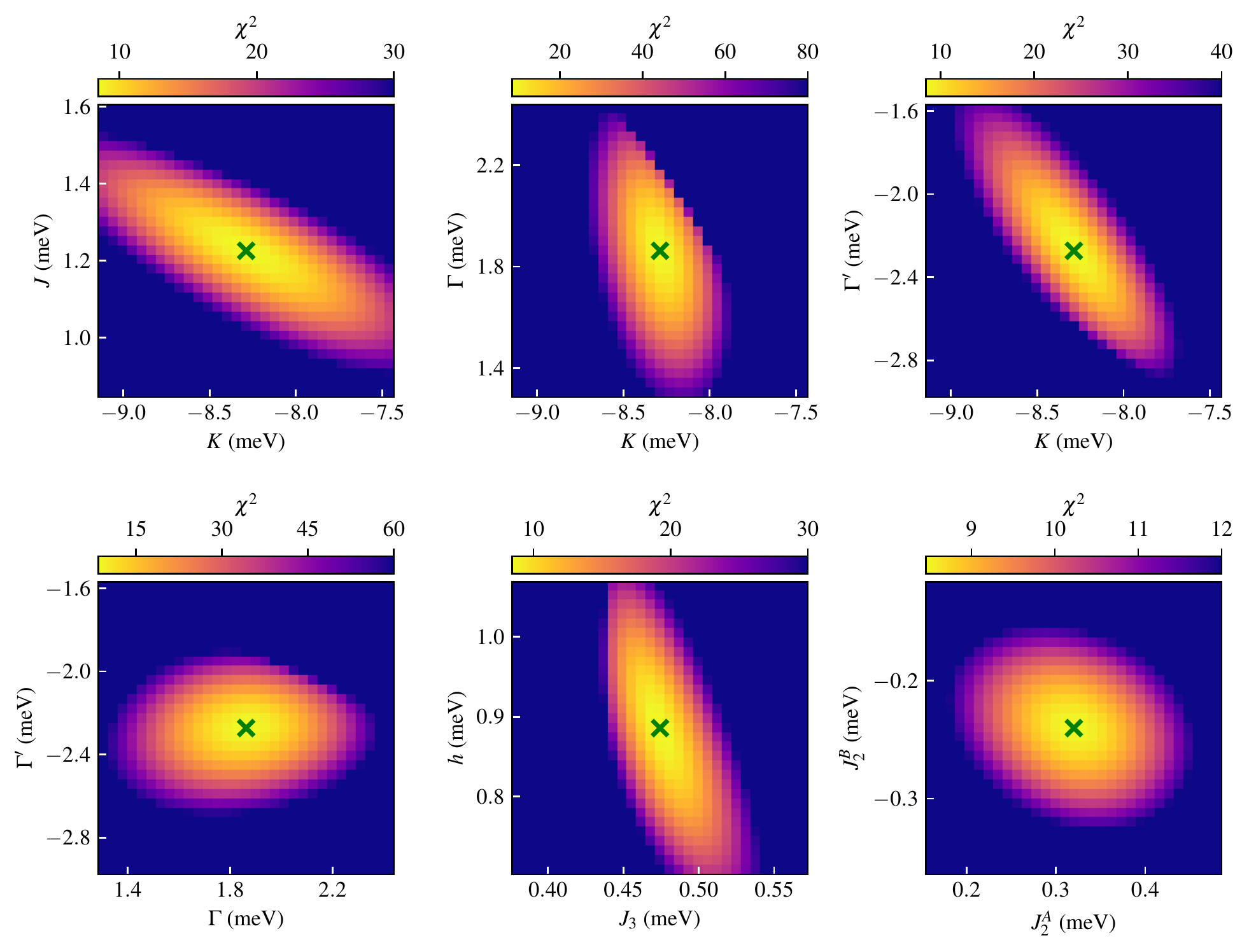}
\caption{Fitness value $\chi^2$ of the HK$\Gamma\Gamma'$ model in several two-dimensional parameter planes. The green crosses correspond to the parameters of the best-fit model mentioned in the main text.}
\label{fig:chisquared_supp}
\end{figure*}

\begin{table*}[tbp]
    \caption{Spin exchange interactions in meV in best-fit HK$\Gamma\Gamma'$ models of \NCTO\ for $S=1/2$. The couplings have been optimized assuming a triple-$\mathbf q$ ground state (first two rows) and a zigzag ground state (third row), respectively.}
    \begin{ruledtabular}\label{tab:best-fit_models}
        \begin{tabular*}{\textwidth}{lcccccccc}
            \quad & $J_1$ & $K_1$ & $\Gamma_1$ & $\Gamma_1^\prime$ & $J_2^A$ & $J_2^B$ & $J_3$ & Nonbilinear interaction\\
            \colrule
            Best-fit triple-$\mathbf q$ model with local-field term\footnote{This is the best-fit model referred to in the main text.} & 1.23 & -8.29 & 1.86 & -2.27 & 0.32 & -0.24 & 0.47 & $h = 0.88$\\
            Best-fit triple-$\mathbf q$ model with ring exchange & 0.68 & -7.89 & 3.07 & -2.94 & -0.06 & -0.70 & 0.52 & $J_{\hexagon}S^4 = -0.26$\\
            Best-fit zigzag model & 1.31 & -8.01 & 0.95 & 0.39 & -0.60 & 0.32 & 1.41 & n/a
        \end{tabular*}
    \end{ruledtabular}
\end{table*}

Three representative spin-wave dispersions, which have the strongest scattering intensities, have been chosen as fitting targets.
They consist of one dispersive band at $\sim$ 1 to 3 meV (mode \#1), and two flat bands at $\sim$ 6 meV (mode \#3) and $\sim$ 7 meV (mode \#4).
For an efficient fitting procedure, we extract the spin-wave dispersions from the neutron scattering intensity using phenomenological functions $E_\text{exp}^{\#\beta}(\mathbf k)$ put forward previously in Ref.~\cite{yao22a},
\begin{equation}
	E_\text{exp}^{\#\beta}(\mathbf k) = 3 J^{\#\beta}_3 S \sqrt{\left( 1 + \frac{2\Delta^{\#\beta}}{3 J^{\#\beta}_3} \right) - |\gamma_{\mathbf k}|^2}\,,
\end{equation}
where $\beta = 1, 3, 4$ corresponds to the targeted spin-wave mode and $\gamma_{\mathbf k} = \frac13 \sum_{\boldsymbol \delta} \rme^{\rmi \mathbf k \cdot \boldsymbol\delta}$, with $\boldsymbol\delta$ the three nearest-neighbor vectors on the honeycomb lattice.
For a given single spin-wave mode~$\#\beta$, the above function can be understood as magnon dispersion of an anisotropic third-neighbor Heisenberg model with energy scale $J^{\#\beta}_3$ and anisotropy parameter~$\Delta^{\#\beta}$.
We emphasize, however, that different bands require different values of $J^{\#\beta}_3$ and $\Delta^{\#\beta}$.
Here, we aim at finding a closed model that describes \emph{all} bands in the magnetic excitation spectrum simultaneously, using a single optimized 8-dimensional parameter set.
The extracted spin-wave dispersions for the targeted bands using the best-fit values for $J^{\#\beta}_3$ and $\Delta^{\#\beta}$~\cite{yao22a} are shown along the momentum trajectory  $(0,0) \rightarrow (0.5,0) \rightarrow (-0.5,0.5)$ in Fig.~\ref{fig:GA_refine_comparison_supp}(c,d) as gray dash-dotted curves. 
Note that along this trajectory, the dispersions are the same as along $\boldsymbol{\Gamma}(0,0) \rightarrow \mathbf{M}(0.5,0) \rightarrow \mathbf{K}(1/3,1/3) \rightarrow \boldsymbol{\Gamma}(0,0)$.
The phenomenological functions $E_\text{exp}^{\#\beta}(\mathbf k)$ are evaluated at totally 200 momentum points (100 for each straight path), to be compared with the calculated spin-wave dispersions of the full HK$\Gamma\Gamma'$ model.

To define a fitness value $\chi^2$ that allows us to quantify the agreement between experiment and model calculation, we first need to designate the eight calculated modes to the three selected experimental targets \#1, \#3, and \#4.
The calculated mode with the lowest energy is designated to the experimental mode with the lowest energy (mode \#1).
The other two experimental modes (modes \#3 and \#4) are then compared with the closest calculated modes in energy.
Fitting with this strategy, we have found that the lowest two branches always stuck together in all satisfying models, which indicates that they are degenerate or quasi-degenerate modes.
Therefore, we have updated our fitting strategy to compare the two lowest-energy modes in calculation with the lowest-energy mode in experiments.
The fitness value of a given parameter set $(J_1, K_1, \Gamma_1, \Gamma'_1, J_2^A, J_2^B, J_3, h)$ of the HK$\Gamma\Gamma'$ model is then defined as
\begin{align}\label{eq:chi2}
	\chi^2 =& \sum_{\mathbf{k}}\left[E_{\text{HK$\Gamma\Gamma'$}}^{\#1}(\mathbf{k}) - E_{\text{exp}}^{\#1}(\mathbf{k})\right]^2 
	\nonumber \\ &
	+ \sum_{\mathbf{k}}\left[E_{\text{HK$\Gamma\Gamma'$}}^{\#2}(\mathbf{k}) - E_{\text{exp}}^{\#1}(\mathbf{k})\right]^2
	\nonumber \\ &
	+ \sum_{\mathbf{k}}\left[E_{\text{HK$\Gamma\Gamma'$}}^{\#a}(\mathbf{k}) - E_{\text{exp}}^{\#3}(\mathbf{k})\right]^2 
	\nonumber \\ &
	+ \sum_{\mathbf{k}}\left[E_{\text{HK$\Gamma\Gamma'$}}^{\#b}(\mathbf{k}) - E_{\text{exp}}^{\#4}(\mathbf{k})\right]^2,
\end{align}
where $E_{\text{HK$\Gamma\Gamma'$}}^{\#\alpha}(\mathbf{k})~(\alpha = 1,2,3,\dots,8)$ is the spin-wave energy at momentum $\mathbf{k}$ of mode $\#\alpha$ from the HK$\Gamma\Gamma'$ model in linear spin-wave theory, $E_{\text{exp}}^{\#\beta}(\mathbf{k})~(\beta = 1,3,4)$ is the spin-wave energy at momentum $\mathbf{k}$ of mode $\#\beta$ from experiment.
Squared differences are summed over 200 momentum points on the trajectory $(0,0) \rightarrow (0.5,0) \rightarrow (-0.5,0.5)$.
Modes $\# a$ and $\# b$ with $a,b \in\{3,4,5,6,7,8\}$, $a\neq b$, are automatically determined such that the last two terms in Eq.~\eqref{eq:chi2} acquire the lowest values.

Although we do not need to specify the boundaries of all parameters in the genetic algorithm, we have started the fitting by limiting the eight parameters in the following zones to speed up the fitting process:
First-neighbor interactions $J_1, K_1, \Gamma_1, \Gamma'_1$ can take arbitrary values in $[-15,15]$~meV.
Second-neighbor interactions $J_2^A, J_2^B$ and third-neighbor interaction $J_3$ can take arbitrary values in $[-5,5]$~meV.
Finally, the strength $h$ of the local-field term can take an arbitrary value in $[0, 5]$~meV.
The boundaries have been refined during the fitting process by examining the fitting result.

Figure~\ref{fig:GA_refine_comparison_supp} demonstrates one complete fitting process of the spin-wave model for \NCTO\ in a constrained 8-dimensional parameter space.
In the first step, we have used the genetic algorithm to find the approximate location of the global minimum.
We have found that 50 generations were enough for the genetic algorithm to complete its task: although the best fitness value slightly dropped after evolving another 150 generations, it converged to the same best-fit model after a local optimization.
The local optimization ensures that the best-fit model has the lowest $\chi^2$ in the neighborhood, see Fig.~\ref{fig:chisquared_supp} for the $\chi^2$ evaluated in several two-dimensional parameter planes.
The spin-wave dispersions of the best-fit model from the last generation of the genetic algorithm and after a local optimization are shown in Fig.~\ref{fig:GA_refine_comparison_supp}(c) and (d), respectively.

\begin{figure*}[tbp]
\includegraphics[width=\linewidth]{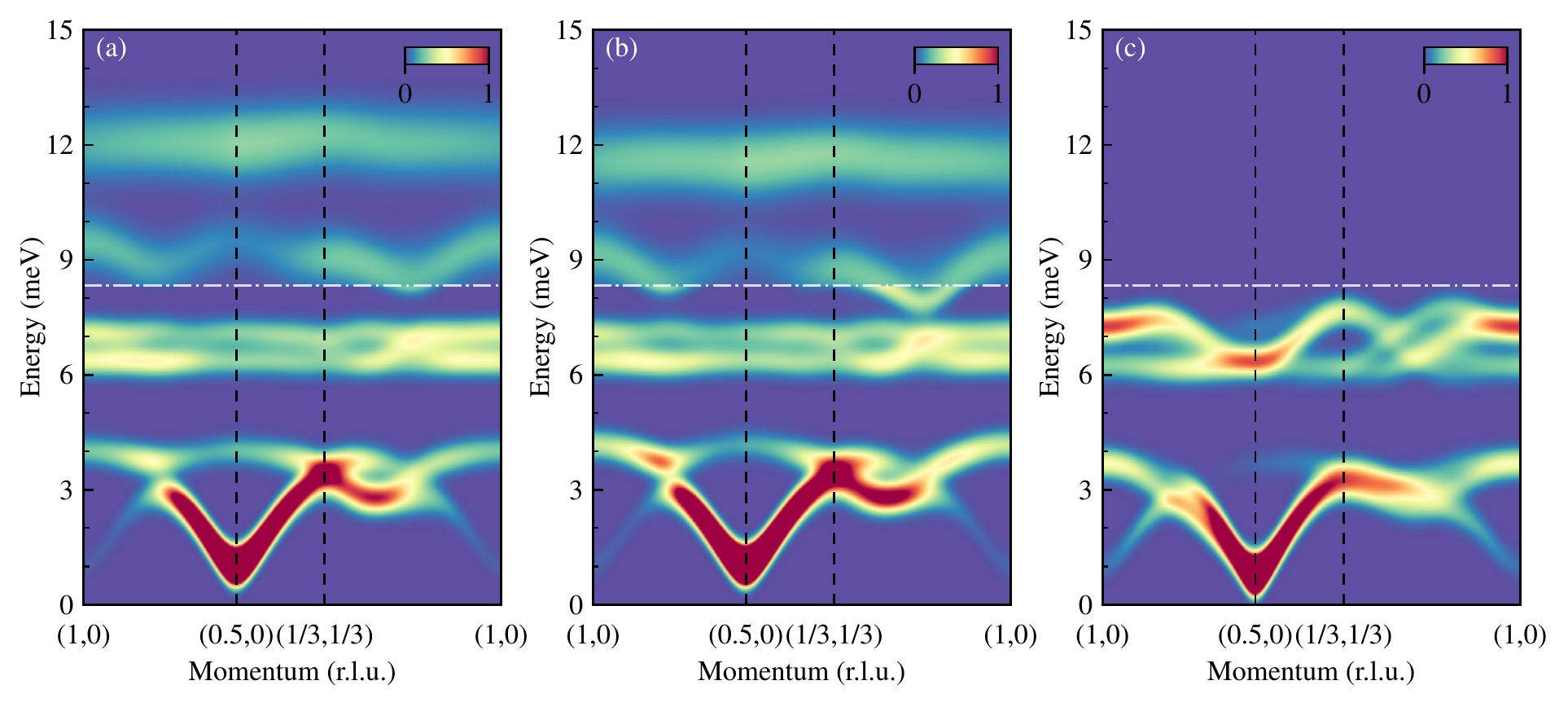}
\caption{%
(a) Magnetic excitation spectrum from linear spin-wave theory of best-fit triple-$\mathbf q$ model with local-field term, with parameters listed in first row of Table~\ref{tab:best-fit_models} (same as Fig.~\figspectrumHmodel\ in the main text).
(b) Same as (a), but for the best-fit triple-$\mathbf q$ model with ring exchange, with parameters listed in the second row of Table~\ref{tab:best-fit_models}.
(c) Same as (a), but for the best-fit zigzag model, with parameters listed in the third row of Table~\ref{tab:best-fit_models}.}
\label{fig:spectrum-LvsHmodel}
\end{figure*}

\begin{figure*}[tbp]
\includegraphics[width=\linewidth]{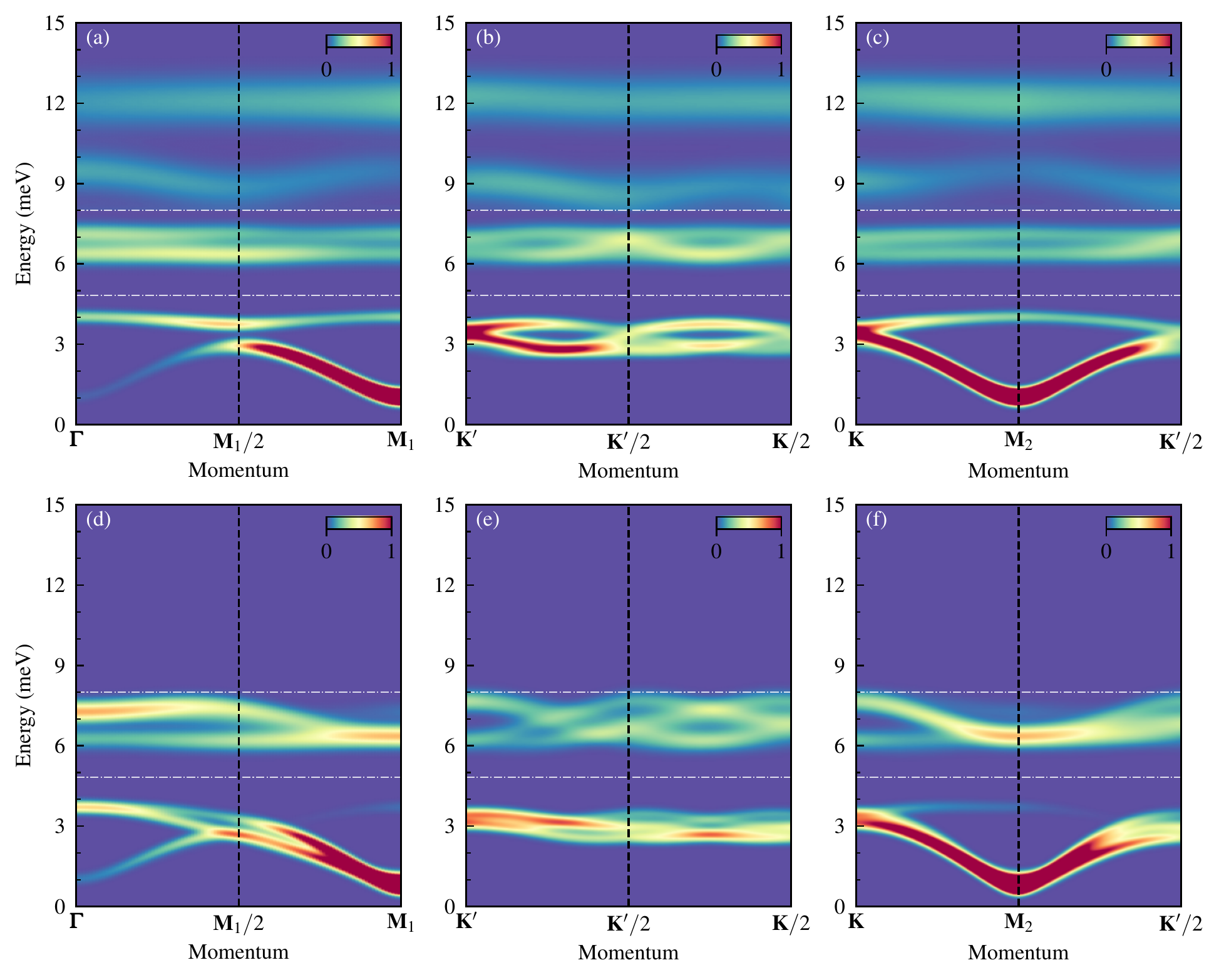}
\caption{(a--c) Magnetic excitation spectrum from linear spin-wave theory of best-fit triple-$\mathbf q$ model with local-field term along momentum trajectories indicated in insets of Figs.~\figspectrumSUtwo(b--g) of the main text, to be compared with Fig.~\figspectrumNCTO\ of the main text. Different full width at half maxima (FWHM) are used in simulations at different energy ranges for better comparison with experimental data: FWHM($E>8.0~\text{meV}$) = 1.18 meV, FWHM($4.8~\text{meV}<E<8.3~\text{meV}$) = 0.59 meV, FWHM($E<4.8~\text{meV}$) = 0.35 meV.
The magnon bands are fully symmetric along these paths as a consequence of the triple-$\mathbf q$ ground state, in agreement with the experimental result for \NCTO, see Fig.~\figspectrumNCTO\ of the main text.
(d--f) Same as (a--c), but for the best-fit zigzag model, averaged over all three zigzag domains. For a zigzag ground state, the magnon bands are not symmetric along these paths, in contrast to the experimental result.}
\label{fig:spectrum-Hmodel_3trajectories_supp}
\end{figure*}

Using the same method, we have also optimized a spin-wave model with higher-order spin exchange directly.
For this, we have used the ring exchange of Eq.~(\eqring) of the main text, parameterized by the coupling $J_{\hexagon}$, instead of the effective local-field term parameterized by $h$.
For simplicity, we have here fixed the spin ordering to realize the triple-$\mathbf q$ order given in Eq.~\eqref{eq:triple-Q}, and omit the optimization of the ground state, in contrast to our procedure above.
As a consequence, the resulting best-fit model in this case does not necessarily have a triple-$\mathbf q$ ground state. We have verified, however, that the triple-$\mathbf q$ state around which we expand corresponds to at least a local energy minimum, leading to a real spin-wave spectrum.
The parameters for the best-fit model with the ring exchange are shown in Table~\ref{tab:best-fit_models} along with the parameters for the best-fit model with the local-field term [Eq.~(\eqfit) of the main text] for comparison.
The bilinear terms for these two models are expected to be qualitatively similar and quantitatively different, since the linear spin-wave theory of the ring exchange involves multiple effective bilinear terms, resulting in renormalizations of the original bilinear terms in the spin-wave model.
We emphasize that different symmetry-allowed nonbilinear terms lead to different such renormalizations, which is why we have constrained ourselves to the local-field term in the fitting discussed in the main text.

Figures~\ref{fig:spectrum-LvsHmodel}(a--b) compares the spin-wave spectra of these two best-fit triple-$\mathbf q$ models, including averaging over triple-$\mathbf q$ domains related by time reversal.
The two spin-wave spectra both match the experimental data very well.
Note that in contrast to the triple-$\mathbf q$ ground state of the best-fit model with local-field term (first row of Table~\ref{tab:best-fit_models}), the actual ground state of the best-fit model with ring exchange (second row of Table~\ref{tab:best-fit_models}) turns out to feature single-$\mathbf q$ order. We emphasize, however, that the good match between experimental and computed magnetic excitation spectra is obtained only for a spin-wave expansion around the triple-$\mathbf q$ state, further substantiating our claim that \NCTO\ features triple-$\mathbf q$ order.

Although a general genetic algorithm does not guarantee to find the global minimum, our best-fit model is likely to reach the global minimum in the constrained parameter space, since multiple independent runs converged to the same best-fit model.
Different initial populations have been generated in these runs, resulting in different optimization routes.
We have also tried different population sizes in genetic algorithms, and they all consistently find the same best-fit model as long as they reach a sufficiently large population size.

We note that for a given parameter set of the HK$\Gamma\Gamma'$ model, a corresponding parameter set that produces the same spin-wave spectrum can always be obtained by a $\mathcal{T}_1$ transformation~\cite{sanders22}.
Since $\mathcal{T}_1$ corresponds to a global $\pi$ rotation around the $[111]$ axis, it will leave all Heisenberg couplings ($J_2^A,J_2^B,J_3$) unchanged, and map the first-neighbor interactions as~\cite{chaloupka15}
\begin{equation}
    \begin{pmatrix}
    J_1\\
    K_1\\
    \Gamma_1\\
    \Gamma'_1
    \end{pmatrix}
    \mapsto
    \begin{pmatrix}
     1 & +\frac{4}{9} & -\frac{4}{9} & +\frac{4}{9}\\
     0 & -\frac{1}{3} & +\frac{4}{3} & -\frac{4}{3}\\
     0 & +\frac{4}{9} & +\frac{5}{9} & +\frac{4}{9}\\
     0 & -\frac{2}{9} & +\frac{2}{9} & +\frac{7}{9}
    \end{pmatrix}
    \begin{pmatrix}
    J_1\\
    K_1\\
    \Gamma_1\\
    \Gamma'_1
    \end{pmatrix}.
\end{equation}
Such dual model of our best-fit triple-$\mathbf q$ model with local-field term is
\begin{align} \label{eq:fit_dual}
& (J_1, K_1, \Gamma_1, \Gamma'_1, J_2^A, J_2^B, J_3)_\text{fit-dual} = 
    \nonumber \\ & \quad
(-4.30, 8.28, -3.66, 0.49, 0.32, -0.24, 0.47)\,\text{meV}\,,
\end{align}
and $h = 0.88~\text{meV}$, for $S = 1/2$. The local fields are now along the $\mathcal{T}_1$-transformed directions.
This alternative model is proximate to a dual version of the hidden-SU(2)-symmeytric point defined in Eq.~(\eqSUtwo), located at $(J_1, K_1, \Gamma_1, \Gamma'_1, J_2)_\text{SU(2)'} = (-1, 2, 0, 0, 0)A$.
Although it has exactly the same spin-wave spectrum as our best-fit model, it features a large positive $K_1$ term, which is not suggested from microscopics~\cite{liu18,sano18}. Therefore, we choose our best-fit model as in Eq.~(\eqfit) of the main text.

We emphasize that while other hidden-SU(2)-symmetric points exist in parameter space~\cite{chaloupka15}, these are obtained from duality transformations with different sublattice structures, and as such feature ordering patterns that are inconsistent with the Bragg peaks observed in \NCTO~\cite{chen21}. The hidden-SU(2)-symmetric point discussed in the main text as a starting point to understand the magnetic properties of \NCTO\ is thus unique.

It is instructive to compare the above results, which assumed triple-$\mathbf q$ order, with those of a model featuring zigzag order. For this, we have optimized a spin-wave model with bilinear interactions only, leading to a zigzag ground state in the classical limit. 
As the domain-averaged zigzag states feature 12 branches of spin waves, which usually run across each other, the branch-to-branch fitting method used above is not possible in this case. As an alternative fitting procedure, we therefore use the intensity information rather than the dispersion, which leads to the parameters denoted as ``best-fit zigzag model'' in Table~\ref{tab:best-fit_models}. The corresponding spin-wave spectrum, including  averaging over the three different zigzag domains, is shown in Fig.~\ref{fig:spectrum-LvsHmodel}(c). We observe that the zigzag model does not reproduce well the flat bands seen experimentally at 6 and 7 meV, respectively (denoted as No.~3 and No.~4, respectively, in Ref.~\cite{yao22a}). Moreover, as all three zigzag domains have the same spectrum at zero momentum, leading to at most four different magnon energies at the $\Gamma$ point in the Brillouin zone, the high-energy features seen experimentally at 9 and 12 meV, respectively (denoted as No.~5 and No.~6, respectively, in Ref.~\cite{yao22a}) in the experiment would need to be necessarily interpreted as multimagnon contributions, which seems somewhat unnatural considering their rich variety of dynamic structure factors at the different energies, as well as their locations on the energy axis.
Most importantly, however, the zigzag model does not reproduce the symmetry of the excitation spectrum, as we discuss in the next supplemental section.

\section{Comparison between theory and experiment}

In this supplemental section, we show more comparisons between theory and experiment.
Figures~\ref{fig:spectrum-Hmodel_3trajectories_supp}(a--c) show the calculated magnetic excitation spectra of the best-fit model mentioned in the main text (first row of Table~\ref{tab:best-fit_models}) along the same momentum trajectories as in Fig.~\figspectrumNCTO\ of the main text.
The calculated spectra are symmetric with respect to these paths, which is guaranteed by the symmetry of the triple-$\mathbf q$ order.
For comparison, Figs.~\ref{fig:spectrum-Hmodel_3trajectories_supp}(d--f) show the corresponding results for the best-fit zigzag model (third row of Table~\ref{tab:best-fit_models}).
Despite the fitting, the domain-averaged spectra are clearly not symmetric along these momentum trajectories, in contrast to the experimental results shown in Fig.~\figspectrumNCTO\ of the main text.

\begin{figure*}[tpb]
\includegraphics[width=\linewidth]{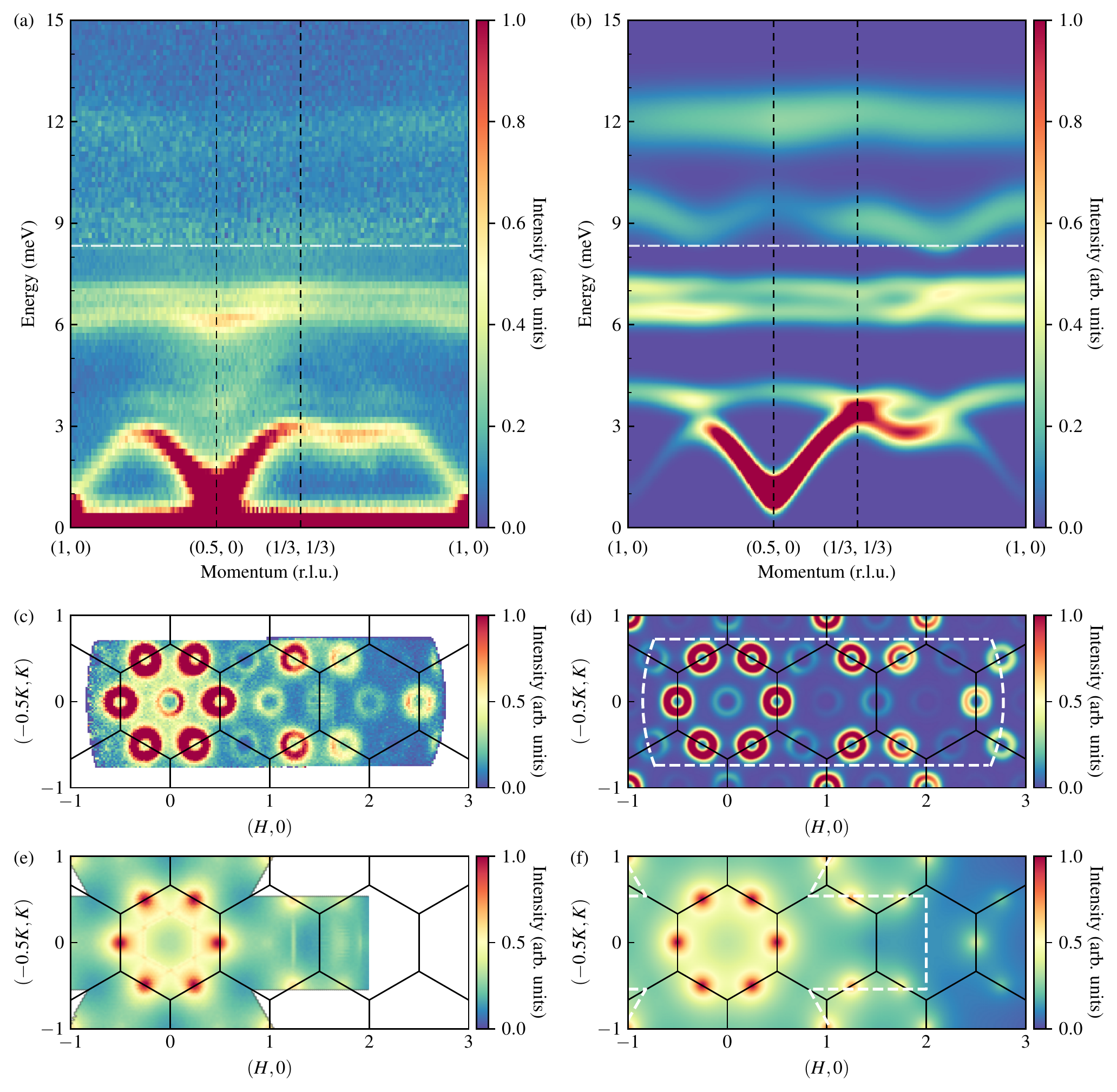}
\caption{Side-by-side comparison between experimental data from neutron scattering of \NCTO\ at $T = 5~\mathrm{K}$ (left) and linear spin-wave theory of best-fit triple-$\mathbf q$ model with local-field term (right). (a,b) Magnetic excitation spectrum along momentum trajectory in $(H,K)$ plane: $(1,0) \rightarrow (0.5,0) \rightarrow (1/3,1/3) \rightarrow (1,0)$. (c,d) Magnetic excitation spectrum integrated over energies $E \in [1.8,~2.2]~\text{meV}$, which is an energy cut through the lowest spin-wave dispersion (mode \#1). (e,f) Magnetic excitation spectrum integrated over large energy range $E \in [0.5,14.2]~\text{meV}$, which can be viewed as equal-time structure factor. White regions indicate incomplete data in the energy range $E \in [0.5,14.2]~\text{meV}$. In (e), the experimental data has been symmetrized by crystallographic point group symmetry operations for easier comparison with the corresponding theoretical data in~(f). The straight-line signals connecting $\mathbf{M}$ points in (e) arise from measurements artifacts at low energy.}
\label{fig:spectrum_side-by-side_supp}
\end{figure*}

Figure~\ref{fig:spectrum_side-by-side_supp} shows a side-by-side comparison between experimental data from neutron scattering at $T = 5~\mathrm{K}$ (left) and linear spin-wave theory of the best-fit model mentioned in the main text, i.e., with triple-$\mathbf q$ ground state (right).
Here, Figs.~\ref{fig:spectrum_side-by-side_supp}(a,c) are generated using the same raw data as those behind Figs.~3(a,h) of Ref.~\cite{yao22a}.
Figure~\ref{fig:spectrum_side-by-side_supp}(b) is the calculated spin-wave spectrum of the best-fit model, which is the same as Fig.~\figspectrumHmodel\ of the main text.
Although the fitting target only includes three modes, all the other modes automatically appear at the correct energies.
This ``coincidence'' strongly indicates that our best-fit model correctly reproduces the main features of the energy landscape near the ground-state spin structure of \NCTO, which is hardly possible if the spin-wave calculation used an incorrect ordered state around which to expand.

\begin{table*}[t]
\caption{Minimal HK$\Gamma\Gamma'$ models for different Kitaev materials, used for exact diagonalization results shown in Fig.~\ref{fig:ed}. The model for \NCTO\ includes an additional nonbilinear coupling as discussed in the main text. The critical value $\alpha_\mathrm{c}$ given in the last column marks the transition between the Kitaev spin liquid and the long-range-order phase, and can be understood as a measure of proximity of the material to the Kitaev regime, with a larger (smaller) value of $\alpha_\mathrm{c}$ corresponding to the respective material being closer to (further away from) the Kitaev spin liquid phase.}
\begin{ruledtabular}
\begin{tabular*}{\textwidth}{lccccccccc}
Material & Reference & \multicolumn{1}{c}{$\mathcal H_\text{Kitaev}$} & \multicolumn{6}{c}{$\mathcal H_{\mathrm{H}\Gamma \Gamma'}$} & $\alpha_\mathrm{c}$ \\
&& $K_1$ (meV) & $J_1$ (meV) & $\Gamma_1$ (meV) & $\Gamma_1'$ (meV) & $J_2^A$ (meV) & $J_2^B$ (meV) & $J_3$ (meV) \\
\colrule
\NCTO & this work & -8.29 & 1.23 & 1.86 & -2.27 & 0.32 & -0.24 & 0.47 & 0.135\\
Na$_2$IrO$_3$ & \cite{winter16} & -17.00 & -- & -- & -- & -- & -- & 6.80 & 0.086 \\
$\alpha$-RuCl$_3$ & \cite{winter17} & -5.00 & -0.50 & 2.50 & -- & -- & -- & 0.50 & 0.175
\end{tabular*}
\end{ruledtabular}
\label{tab:ed}
\end{table*}

\begin{figure*}[t]
\includegraphics[width=\textwidth]{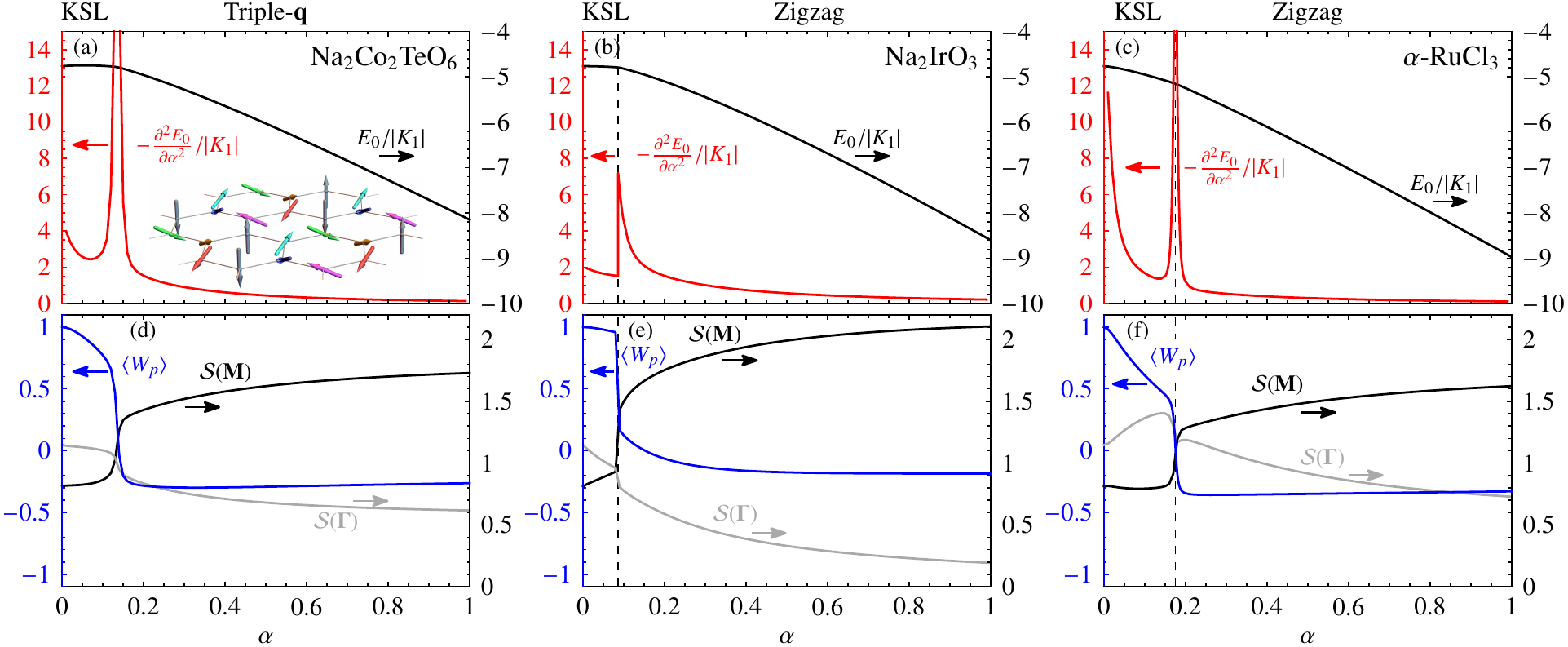}
\caption{Top row: Ground-state energy $E_0$ and second derivative $-\partial^2E_0/(\partial \alpha^2)$ as function of the parameter $\alpha$, which interpolates between the pure Kitaev limit for $\alpha = 0$ and minimal spin models for (a) \NCTO, (b) Na$_2$IrO$_3$, and (c) $\alpha$-RuCl$_3$ for $\alpha = 1$, from exact diagonalization on a 24-site hexagonal cluster. The dashed lines indicate the transitions from the Kitaev spin liquid (KSL) to (a) the triple-$\mathbf q$ and (b,c) the zigzag phases, respectively. The inset in (a) shows the local magnetization for the best-fit \NCTO\ model for $\alpha = 1$, illustrating the triple-$\mathbf q$ ground state. Bottom row: Same as (a,b,c), but showing the plaquette expectation value $\langle W_p \rangle$ and the static structure factor $\mathcal S(\mathbf q)$ at the $\boldsymbol\Gamma$ and $\mathbf M$ points of the first Brillouin zone.}
\label{fig:ed}
\end{figure*}

Figure~\ref{fig:spectrum_side-by-side_supp}(d) shows the theoretical spectrum integrated over energies $E \in [1.8, 2.2]~\text{meV}$, to be compared with the corresponding experimental data in Fig.~\ref{fig:spectrum_side-by-side_supp}(c).
The relative intensities at different momenta match the experimental data very well.
We emphasize again that the fitting target consists of only three dispersions along one momentum trajectory without any intensity information.
Therefore, the remarkably good agreement between theory and experiment in terms of intensity renders our best-fit triple-$\mathbf q$ model a successful spin-wave model of \NCTO.
Comparing Figs.~\ref{fig:spectrum_side-by-side_supp}(c) and (d) carefully, one may notice that the intensities at the $\boldsymbol{\Gamma}$ points [i.e.,  $\mathbf{q}  = (0,0),(1,0),(2,0),\dots$] are relatively stronger in experimental data.
We infer that this difference may come from magnetic excitations beyond linear spin-wave theory, such as fractional excitations.
We leave this intriguing signature of possible fractional excitations for future work.

Finally, Figs.~\ref{fig:spectrum_side-by-side_supp}(e,f) show the spectrum integrated over a large energy range $E \in [0.5,14.2]~\text{meV}$ from \NCTO\ and best-fit model, respectively. Such energy-integrated inelastic signal can be viewed as equal-time structure factor~\cite{plumb16}, and depends on the ``snapshot'' of the ground-state spin structure. The good agreement between integrated experimental data and integrated theoretical spectrum indicates that they both feature the same triple-$\mathbf q$ ground state.

\begin{figure*}[bt]
\includegraphics[width=\linewidth]{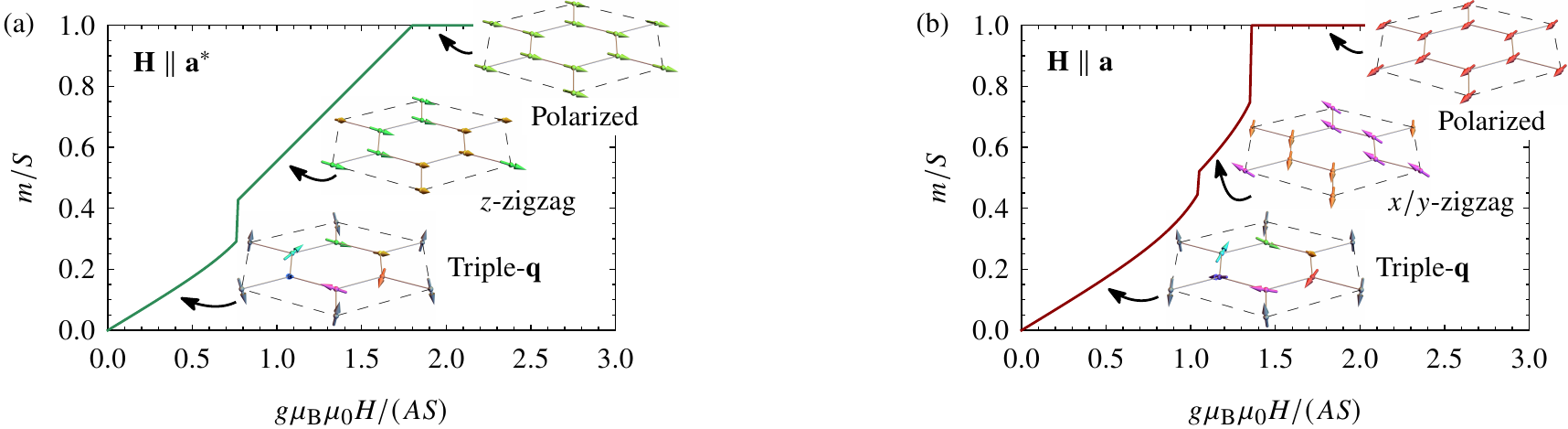}
\caption{(a) Classical magnetization curve in HK$\Gamma\Gamma'$ model with bilinear exchange couplings
$(J_1,K_1,\Gamma_1,\Gamma'_1,J_2^A, J_2^B,J_3)S^2/A = (-1/9,-2/3,8/9,-4/9,0,0,0)$ and small ring exchange coupling $J_{\hexagon}S^6/A = -1/10$ as function of in-plane field along the armchair direction, $\mathbf H \parallel \mathbf a^*$, showing a metamagnetic transition between the triple-$\mathbf q$ state at low fields and the canted $z$-zigzag state at intermediate fields. The insets illustrate the spin configurations for representative values of $H$. (b) Same as (a), but for an in-plane field along the zigzag direction, $\mathbf H \parallel \mathbf a$, with a metamagnetic transition between the triple-$\mathbf q$ state and canted $x$- and $y$-zigzag states.}
\label{fig:magnetization}
\end{figure*}

\section{Proximity to Kitaev quantum spin liquid}

In this supplemental section, we discuss the proximity of our best-fit model for $S=1/2$ to the Kitaev quantum spin liquid, using exact diagonalization on a 24-site hexagonal cluster.
To this end, we consider a one-dimensional family of extended Heisenberg-Kitaev-Gamma-Gamma$'$ models involving the parameter $\alpha$, which controls the proximity of the model to the pure Kitaev limit. In particular, we study the ground-state phase diagram of the family of Hamiltonians
\begin{align}
    \mathcal H(\alpha) = \mathcal H_\text{Kitaev} + \alpha \mathcal H_{\mathrm{H}\Gamma \Gamma'},
\end{align}
where $\mathcal H_\text{Kitaev} = K_1 \sum_{\langle ij\rangle} S_i^\gamma S_j^\gamma$ denotes the Kitaev interaction, and $\mathcal H_{\mathrm{H}\Gamma\Gamma'}$ includes all other bilinear and nonbilinear exchange interactions, such that $\mathcal H(\alpha)$ interpolates between the pure Kitaev limit for $\alpha = 0$, i.e., $\mathcal H(0) = \mathcal H_\text{Kitaev}$, and the best-fit model for \NCTO\ for $\alpha = 1$, i.e., $\mathcal H(1) = \mathcal H_\text{\NCTO}$.
We compare our results for \NCTO\ with corresponding results using analogously defined families of models that interpolate between the pure Kitaev model and accepted minimal models for the Kitaev candidate materials Na$_2$IrO$_3$ and $\alpha$-RuCl$_3$.
The corresponding parameter sets are given in Table~\ref{tab:ed}.
Note that the family of models describing Na$_2$IrO$_3$ and $\alpha$-RuCl$_3$ for $\alpha = 1$ feature single-$\mathbf q$ zigzag ground states, in contrast to the triple-$\mathbf q$ ground state of our best-fit model for \NCTO. We emphasize that each family of models represents a different direction away from the pure Kitaev limit in the multidimensional parameter space. The chosen models for Na$_2$IrO$_3$ and $\alpha$-RuCl$_3$ in Table~\ref{tab:ed} are according to the current literature.
To characterize the ground state as function of $\alpha$, we compute the ground-state energy and the expectation value of the plaquette operator,
\begin{align}
    W_p = 2^6 S^x_1 S^y_2 S^z_3 S^x_4 S^y_5 S^z_6,
\end{align}
where the indices $1,2,\dots,6$ denote the six sites of an elemental plaquette in clock-wise order, with $\langle 12\rangle$, $\langle 23\rangle$, and $\langle 34 \rangle$ corresponding to $z$, $x$, and $y$ bonds, respectively. The plaquette expectation value approaches unity in the pure Kitaev limit~\cite{kitaev06} and becomes negative for a semiclassical zigzag state~\cite{gordon19}.
We also compute the static structure factor defined as
\begin{align}
    \mathcal S(\mathbf q) = \frac{1}{N} \sum_{i,j} \langle \mathbf S_i \cdot \mathbf S_j \rangle \rme^{\rmi \mathbf q \cdot (\mathbf R_i - \mathbf R_j)} ,
\end{align}
where $\mathbf R_i$ is the lattice vector at site $i$ and $N = 24$ the total number of sites.
For all three families of models, each observable shows a distinguished anomaly at a unique critical value $\alpha_\mathrm{c}$, indicating a direct quantum phase transition between the paramagnetic Kitaev quantum spin liquid for $\alpha < \alpha_\mathrm{c}$ and an antiferromagnetic long-range-ordered phase for $\alpha > \alpha_\mathrm{c}$, see Fig.~\ref{fig:ed}.
For the family of models relevant to \NCTO, the transition is towards a triple-$\mathbf q$ state, as illustrated by the local magnetization for $\alpha = 1$ shown in the inset of Fig.~\ref{fig:ed}(a).
For the families of models relevant to Na$_2$IrO$_3$ and $\alpha$-RuCl$_3$, the transition is towards a single-$\mathbf q$ zigzag ground state~\cite{winter16, winter17b}.
Importantly, since $\alpha$ interpolates between the pure Kitaev limit for $\alpha = 0$ and the minimal spin model for a given material for $\alpha = 1$, the corresponding material-dependent critical value $\alpha_\mathrm{c}$ can be understood as a measure of proximity of the material to the Kitaev quantum spin liquid regime.
We find that $\alpha_\mathrm{c}(\text{$\alpha$-RuCl$_3$}) > \alpha_\mathrm{c}(\text{\NCTO}) > \alpha_\mathrm{c}(\text{Na$_2$IrO$_3$})$, suggesting that \NCTO\ is not as close to the Kitaev spin liquid as $\alpha$-RuCl$_3$, but closer than Na$_2$IrO$_3$.
As a side remark, we mention that a recent variational Monte Carlo study finds that a minimal spin model for \NCTO, which is closely related to our best-fit model, and also realizes a triple-$\mathbf q$ ground state, turns out to be proximate to yet another quantum spin liquid phase, suggesting that the realization of a quantum spin liquid in \NCTO\ could in fact be more promising than in other proposed Kitaev candidate materials~\cite{wang23}.
The proximity of \NCTO\ to the spin liquid regimes will be important for the physics in the intermediate temperature regime above the N\'eel temperature, as discussed in the main text. It might be further enhanced by external perturbations, such as magnetic field, strain, or chemical substitution.

\section{Magnetization processes}

In this supplemental section, we demonstrate that a spin model proximate to the hidden-SU(2)-symmetric point generically features a metamagnetic transition between the triple-$\mathbf q$ state at small fields and a canted zigzag state at intermediate fields, before the transition towards the paramagnetic state at high fields.
The phenomenon arises due to the noncollinear ordering in the triple-$\mathbf q$ state already at zero field, implying that many of the spins in the zero-field ground state have sizable components along the field axis, inhibiting an efficient canting of these spins in finite fields. The field therefore favors those canted states that allow for a homogeneous canting, which applies, in particular, to collinear states, such as zigzag states~\cite{janssen17, balz21}.

To be explicit, consider the extended  HK$\Gamma\Gamma'$ model with small ring-exchange perturbation, parametrized by the coupling $J_{\hexagon} < 0$ in Eq.~(\eqring) of the main text, in a finite magnetic field $\mathbf H$, described by the Zeeman term
\begin{align}
\mathcal H_{\mathbf H} = - g \mu_\text{B} \mu_0 \mathbf H \cdot \sum_i \mathbf S_i.
\end{align}
For simplicity, we assume a parameter set for the bilinear exchange couplings corresponding to the hidden-SU(2)-symmetric point, as given in Eq.~(\eqSUtwo) of the main text.
For small ferromagnetic $J_{\hexagon} < 0$, the classical model features a triple-$\mathbf q$ ground state at zero field, see Fig.~\figspectrumSUtwo(a) of the main text.
For external field along the armchair direction, $\mathbf H \parallel \mathbf a^*$, this state does not allow an efficient canting mechanism, as half of the spins in the magnetic unit cell have sizable components along the field axis, see lower inset of Fig.~\figspectrumSUtwo (a) of the main text.
In the $z$-zigzag state, with antiferromagnetic (ferromagnetic) alignment of spins along the $z$ ($x$ and $y$) bonds on the honeycomb lattice, by contrast, all spins in the unit cell are aligned perpendicular to the armchair direction, such that these can cant homogeneously towards the field axis, see upper inset of Fig.~\figspectrumSUtwo (a) of the main text.
For $\mathbf H \parallel \mathbf a^*$ and small ferromagnetic $J_{\hexagon} < 0$, we therefore expect a transition between the triple-$\mathbf q$ state at low fields and the canted $z$-zigzag state at intermediate fields, before the transition towards the high-field state.
This expectation is indeed confirmed by the explicit calculation. Figure~\ref{fig:magnetization}(a) shows the classical magnetization at zero temperature as function of in-plane field along the armchair direction, illustrating the metamagnetic transition that is located roughly halfway between zero external field and the transition towards the high-field phase. In the classical limit, the magnetization curve is linear in the canted zigzag phase, as all spins cant homogeneously towards the field axis for $\mathbf H \parallel \mathbf a^*$.

As similar behavior is found for the zigag in-plane field direction, $\mathbf H \parallel \mathbf a$, see Fig.~\ref{fig:magnetization}(b). The only qualitative difference here is that the magnetization is now nonlinear in both ordered phases, as the spins in neither of the three zigzag states can cant homogeneously towards the $\mathbf a$ direction.

\bibliographystyle{longapsrev4-2}
\bibliography{ncto-duality}